# Anisotropy in Electronic and Magneto-transport of 2D superconductor $NbSe_2$

N. K. Karn[1,2,*], M. M. Sharma[3], I. Felner[4] and V.P.S. Awana[1,2]

[1]Academy of Scientific & Innovative Research (AcSIR), Ghaziabad, 201002, India
[2]CSIR- National Physical Laboratory, New Delhi, 110012, India
[3]Department of Physics, University of Arkansas, Fayetteville, AR 72701, USA
[4]Racah Institute of Physics, The Hebrew University, 91904, Jerusalem, Israel

## Abstract

This article reports the successful synthesis of single crystalline two-dimensional thin flakes of $NbSe_2$. The XRD (X-ray Diffraction) pattern of the grown crystal ensured its crystallization in a single phase with a hexagonal structure. The EDAX (Energy Dispersive X-ray Analysis) endorsed the stoichiometry of the as-grown sample. To study the vibrational modes, the Raman spectra were recorded, which exhibited the expected four Raman active modes. The resistance vs temperature measurement showed a well-established superconducting transition ($T_c$) at 7.3 K. The ZFC (Zero-Field Cooled) & FC (Field Cooled) magnetization curves, as well as the isothermal M−H (Magnetization vs. field) measurements, have been performed for both in-plane and out-of-plane H directions. Distinct anisotropy is observed in both magnetization and magneto-transport measurements with field direction, leading to different critical fields ($H_c$). Out-of-plane magneto-transport data hints towards the existence of a filamentary state. The density functional theory (DFT) has been used to study the band structure of $NbSe_2$. Although the bulk band structure confirmed metallic behavior, the same of mono-layers of $NbSe_2$ within the GGA+U framework showed a band gap of 1.17 eV. The article addresses the anisotropy in the electronic and magneto-transport of 2D superconductor $NbSe_2$.

*Corresponding Author
Mr. N.K. Karn, Senior Research Fellow
CSIR-National Physical Laboratory, India
E-mail: nkk15ms097@gmail.com
Ph. +91-11-45609357, Fax-+91-11-45609310

## Introduction

Superconductivity (SC) in 2D materials has emerged as a captivating frontier in condensed matter physics and materials science in recent years [1]. These materials have led to fruitful physical properties, such as high transition temperatures ($T_c$) [2] and enhanced parallel critical magnetic fields ($H_{c2}$) [3]. 2D SC refers to the phenomenon occurring in extremely thin materials or at the interfaces of materials, whereby the conducting electrons are confined to a two-dimensional plane. Since the advent of nano-fabrication techniques, heterogeneous interfaces have enabled the study of highly crystalline 2D superconductors, unveiling phenomena such as quantum Griffiths singularity [4] and topological superconductivity [5].

Transition metal dichalcogenides (TMDs) are one such 2D class of materials, whose structure contains layers of X-M-X (X: Chalcogenide, M: Transition metal) separated by Van der Waals gaps. $NbSe_2$ is one of the most studied TMDs since its discovery, over some 50 years ago, due to its higher $T_c$ as compared to other TMDs [6-14]. Besides, bulk SC, $NbSe_2$ introduces unique properties and challenges, including proximity effect [7, 8], anisotropy in SC parameters [9], charge density waves [8, 10-13], and Ising superconductivity [14].

In the past decade, topological materials have brought a new paradigm in condensed matter physics. Among these, unlike the topological insulators, the topological semimetals (TSMs) possess not only non-trivial topological conducting surface states but also have conducting bulk states beneath [15]. TMDs belonging to the class of TSMs have attracted the special interest of material scientists, as these materials contain two-dimensional atomic layers interacting via Van der Waals force. They prove to be good candidates for material intercalation, because of their two-dimensional layered structure, which allows the parent material properties to be engineered. For example, $PdTe_2$ is SC with a $T_c$ of 1.7 K [16], which can be increased up to 4.7 K by Au intercalation [17]. $NbSe_2$ is also a TMD with a $T_c$ of 7.2 K [18]. The structural changes due to transition metal substitution [19] and Rb intercalation [20] in $NbSe_2$ have been reported earlier. Mg intercalation shows a nominal effect on $T_c$ of $NbSe_2$ [21]. Finding suitable topological properties such as non-trivial surface states, would make $NbSe_2$ one of the best avenues to realize intrinsic topological superconductivity [22].

Structurally, $NbSe_2$ shows polytypism due to the existence of simultaneous trigonal prismatic (2H) and octahedral (1T) layered crystal structures [23]. Bulk $NbSe_2$ crystallizes into the 2H phase, and the charge density wave and superconductivity emerge simultaneously and interact with each other, thereby leading to various anomalous properties [8]. Its electronic properties make it a suitable candidate for electronic and optoelectronic devices. An atomically thin layer of $NbSe_2$ has shown the capacity to manipulate nonreciprocal sensitivity thereby its application in antenna devices [24]. Although SC properties and their origin in $NbSe_2$ have been well explored, it has still a lot to offer to physicists as demonstrated by recent additions to the literature [25- 27].

Being $NbSe_2$ a layered 2D material, the presence of anisotropy is anticipated and has been explored in terms of its different facets [9, 12, 28-30]. The anisotropic anomaly above $T_c$ in resistivity shows a singularity around 32 K [12]. The angle-dependent resistivity demonstrates the anisotropic properties below $T_c$ [9]. Here we demonstrate the anisotropy presence in $NbSe_2$ by magneto-transport measurements, showing a higher value of the critical field for in-plane as compared to the out-of-plane direction. In this study, the synthesized 2D flakes of $NbSe_2$ are characterized by XRD and Raman spectra. The surface morphology and sample stoichiometry are validated by EDX spectra. The isothermal magnetization (M-H) and M-T measurements are also performed for both directions. Additionally, the band structure of bulk, monolayer and bi-layer of $NbSe_2$ has also been computed and is reported here.

## Experimental and Computational Methodology

There are different solid-state reaction routes to synthesize $NbSe_2$ polycrystals [31] and single crystals [9, 32]. The quality single crystal reported here has been synthesized by the Chemical vapor transport (CVT) method [31]. The pure powders of Nb and Se freshly bought from Alfa Aeser, were mixed in a 1:2 molar ratio. The mixture was made homogenous by thoroughly mixing and grinding them in an inert atmosphere inside an MBraun Glove Box filled with argon gas. This homogenous mixture was then palletized. Thus, the obtained pallet of $NbSe_2$ along with Iodine (transport agent) was vacuum encapsulated in a quartz ampoule under the pressure of $10^{-5}$ mbar. The amount of iodine taken is 5 mg/cm$^3$. The vacuum-encapsulated mixture was placed in a two-zone furnace keeping the hot zone at 800ºC and the cold zone at 700ºC. The sample was dwelled at the mentioned temperature gradient for 14 days and then normally cooled to room temperature. Thin shiny sheets of as synthesized $NbSe_2$ obtained were characterized and electronic and magneto-transport properties were studied. Rigaku Miniflex II tabletop x-ray diffractometer (XRD) equipped with Cu-$K_{\alpha}$ (1.5406 Å) radiation was used for phase purity characterization of synthesized $NbSe_2$ single crystal 2D flakes. The surface morphology and elemental compositions were examined by performing scanning electron microscopy (SEM) and energy-dispersive X-ray analysis (EDAX) measurements using Zeiss EVO-50. Further, to study the vibrational modes Raman spectroscopy was performed. Renishaw in Via Reflex Raman Microscope equipped with a 532 nm laser is used to irradiate the thin flakes of mechanically cleaved $NbSe_2$ crystal to record Raman spectra. To avoid local heating, laser power was maintained below 5 mW during the irradiation process of the flake surface. The conventional four-probe arrangement was used to study the transport properties of synthesized crystals by using a Quantum Design Physical Property Measurement System (QD-PPMS). For the transport measurements, the sample was mounted on a PPMS puck and the electrical contacts were made using Epotek H20E silver epoxy. Temperature-dependent and isothermal magnetization were measured using a Quantum Design SQUID Magnetic Property Measurement System (MPMS).

Theoretically, the band structure of $NbSe_2$ was simulated using the open-source software Quantum Espresso [33, 34]. It works on the first principle-based DFT calculations. The ground state electron density and wavefunction were calculated by solving the Kohn-Sam equation in a self-consistent manner. To include electronic exchange and correlation factor correction in the simulation, Perdew-Burke-Ernzerhof (PBE) type functional was used including generalized gradient approximation (GGA) [35]. In calculation, to take account of Coulomb potential, calculations were performed within the GGA+U framework. The corresponding Hubbard parameter U calculated for the Nb atom is 2.9344 eV. Since the system under investigation is Van der Waals gap material, the Van der Waals force corrections were also implemented using the DFT-D3 module. The atomic optimization for Bulk and bilayer was performed using DFT-D3 correction. The monolayer and bilayer of $NbSe_2$ were modeled by creating a vacuum of ~15 Å above the $NbSe_2$ layers to avoid any extra interlayer interaction. The mono and bi-layers were created using the VESTA software. For the band structure computation, the following parameter setting was used in the input file: The total energy threshold and electronic convergence cut-off were set to $6\times10^{-5}$ and $1.2\times10^{-9}$ Ry respectively. The first Brillouin Zone (BZ) was sampled on a Monkhorst-Pack k-grid of 11×11×3 for the Bulk $NbSe_2$ and the same for monolayer and bilayer was taken to be 11×11×1. For the calculation charge cut-off was set to 344 Ry and wfc was set to 86 Ry.

## Results and Discussion

Fig. 1 displays the XRD pattern of thin flakes obtained from mechanically cleaved $NbSe_2$ single crystal. The pattern shows high-intensity peaks and indicates that the crystal growth has occurred along the c-axis and the diffraction planes are stacked along (002n) where n=1,2,3... The successful synthesis of phase pure crystalline thin sheets of $NbSe_2$ is confirmed by matching the single crystal XRD peaks from the reference [36]. The VESTA-generated unit cell of $NbSe_2$ is shown in the inset of Fig. 1(a). The material under study has been found to show polytypes [23] with Trigonal and two monolayers layered of hexagonal lattices (assigned as 2H-$NbSe_2$). The calculated c lattice parameter is 12.58 ±0.2 Å which is at par with the earlier reports [14, 23].

The phase purity of the synthesized $NbSe_2$ single crystal is verified through Rietveld refinement of powder XRD data and is shown in Fig. 1(b). Rietveld refinement is performed using the Full Prof software. All peaks are well fitted with applied parameters of 2H-$NbSe_2$, which confirms the phase purity of the synthesized $NbSe_2$ single crystal. The quality of fit is determined by $\chi^2$ values, which is found to be 6.51, which lies in an acceptable range. No peaks are found other than $NbSe_2$, showing the absence of any other impurity phase. The atomic coordinates obtained from Rietveld refinement are Nb (0, 0, 0.25) and Se (0.333, 0.666, 0.384). The obtained lattice parameters are a=b= 3.446(5) Å, & c= 12.554(5) Å and $\alpha=\beta=90^o$ & $\gamma=120^o$.

Fig. 2(a) shows the SEM image taken on synthesized single-crystalline 2D flakes of $NbSe_2$. The single crystalline nature of the sample is endorsed by the SEM image, which portrays layered terrace-type morphology. Linear growth along the c-axis is visualized at 2 μm resolution. Also, no other contrasting colour has been observed, which shows that the crystal is grown in a single phase. Fig. 2(b) shows the observed EDAX spectra for the elemental analysis of the crystal, containing the constituent elements Nb and Se only, confirming the absence of any foreign impurity element. The elemental composition is given in the inset table of Fig. 2(b). The observed stoichiometry is $Nb_{1.01}Se_{1.99}$, which is within the uncertainties and is close to $NbSe_2$. Both structural single crystal XRD and microstructural SEM/EDAX characterization approve crystallinity along the (002n) plane, morphology and the stoichiometry of as-grown sample $NbSe_2$.

Further, Raman spectroscopy is used to study the vibrational modes present in the 2D flakes of synthesized $NbSe_2$. TMDs with hexagonal lattice (of the form 2H-$MX_2$) have four Raman active vibrational modes, namely the $A_{1g}$, $E^1{}_{2g}$, $E_{1g}$ and $E^2{}_{2g}$ [37]. Fig. 3 shows recorded Raman spectra of the synthesized $NbSe_2$ single crystal, exhibiting anticipated four Raman active peaks. Observed peaks are fitted, analyzed, and identified through Lorentz deconvolution as shown in Fig 3. Here, mainly four Raman peaks are observed at 94 ± 0.6, 154 ± 0.3, 191 ± 1 and 227 ± 0.2 $cm^{-1}$, which are attributed to low-frequency mode $E^2{}_{2g}$ and high-frequency modes $E_{1g}$, $E^1{}_{2g}$, $A_{1g}$ of $NbSe_2$ crystal, respectively. The results are in agreement with ref. 37.

To investigate the superconducting properties of $NbSe_2$, electronic and magneto-transport measurements have ~~also~~ been performed. Fig. 4 shows the resistivity vs temperature (ρ-T) plot at zero magnetic field. A superconducting transition is observed (see inset of Fig. 4) as the resistivity starts decreasing sharply at $T_c^{onset}$ = 7.3±0.1 K, which is close to the previously reported values [6,18]. The temperature at which ρ =0 ($T_c^{offset}$) is seen, is at7.1±0.1 K, with a transition width of ~0.2 K. Above $T_c$, the resistivity increases with temperature, showing high metallicity of the synthesized crystal. Above $T_c$ (i.e., 7.3–300 K) the ρ–T data are fitted using the Bloch–Grüneisen formula given below

$$(1)\ \rho(T) = \left[\frac{1}{\rho_s} + \frac{1}{\rho_i(T)}\right]^{-1}$$

Here, $\rho_s$ denotes the temperature-independent saturation resistivity [38] and $\rho_i(T)$ is the temperature-dependent term which can be given by:

$$(2)\ \rho_i(T) = \rho(0) + \rho_{el-ph}(T)$$

Here, ρ(0) denotes residual resistivity arising due to impurity scattering and $\rho_{el-ph}(T)$ denotes the temperature-dependent term, which depends on the electron–phonon scattering. Furthermore, $\rho_{el-ph}(T)$ is given by the following formula:

$$(3)\ \rho_{el-ph} = \alpha_{el-ph}\left(\frac{T}{\theta_D}\right)^n \int_0^{\frac{\theta_D}{T}} \frac{x^n}{(1-e^{-x})*(e^x-1)}\,dx$$

Here, $\alpha_{el-ph}$ is the electron–phonon coupling parameter, $\theta_D$ represents the Debye temperature, and n is a constant. The ρ–T data are well fitted for n = 5, signifying dominant electron–phonon scattering. The deduced values are $\theta_D$ =165.85 K and ρ(0) = 20.89 μΩ-cm. The residual resistivity ratio (RRR) given by ρ(300K)/ρ(Tc) is 17.69, which is higher than the previously reported RRR value of 11 [36]. This endorses the quality and the high metallicity of the synthesized crystal.

To illustrate the anisotropic properties of the critical field, the magneto-transport measurements have been performed in two different orientations of the crystal with respect to the applied field (H). Fig. 5(a), (b) and (c) show the ρ-T plots for different H values in the out-of-plane orientation (H || to c). The inset of Fig. 5(a) shows the experimental setup. From Fig. 5(a), it is observed that as the field is increased (say for H>0.1 T), the transitions to the superconducting state take place in two steps. The two-step transition is more apparent in Fig. 5(b), which shows the ρ-T plots H>0.2 T 0.2 T. We define temperature $T_c^*$ as the transition temperature from the normal state to the intermediate state (which is a non-superconducting state). The $T_c^{onset}$ is the transition temperature from the intermediate state or normal state to the superconducting state. The fields corresponding to $T_c^*$ and $T_c^{onset}$ are defined as $H_{c3}$ and $H_{c2}^{onset}$ respectively. For easy notion, these points, ($T_c^{onset}$, $H_{c2}^{onset}$) and ($T_c^*$, $H_{c3}$) are marked in Fig 5 and 6. Fig. 5(d) depicts the ρ-H plot for H || to c at 2 K, 4 K, 6 K and 8 K, where the intermediate state is more pronounced. Such an intermediate state has been reported to be present in superconducting Nb samples [39], showing the possible existence of a filamentary phase. In ref. 38, the formation of filaments parallel to the field irrespective of the field orientation above $H_{c2}$ is reported. In the present study on $NbSe_2$, a similar filamentary phase is observed above $H_{c2}$ for only H || to c, but not for H || to ab. This shows that the formation of the filamentary phase is sensitive to the direction of the applied magnetic field.

Above $T_c$, i.e. at 8K, the normal metallic behavior is observed. The inset of Fig. 5(d) shows the magneto-resistance (MR) % at 8 K. The MR of a usual metal has a quadratic dependence on the applied field. In contrast to that, here the inset shows the linear variation of MR % with the applied magnetic field. This indicates the possibility of the presence of topological surface states [40] in $NbSe_2$, which needs further investigation.

Fig. 6(a) and (b) show the magneto-transport for in-plane field orientation (H ‖ to ab). In this orientation, the superconducting transition is one step only and the intermediate state is not observed.

The anisotropy present in $NbSe_2$ is self-evident by comparing the ρ-T and ρ-H data plots of Fig. 5 and Fig 6. Further, to analyze the anisotropy in the as-synthesized $NbSe_2$, the critical fields ($H_{c2}$) are derived from the magneto-transport data, for both orientations, which are shown in Fig.7. From the previous discussion, the possibility of filamentary phase is elucidated for H ‖ to c. As temperature decreases, the gap $H_{c3}$-$H_{c2}$ increases, conversely the filamentary state becomes more prevalent with H ($\Delta T = T_c^* - T_c^{onset}$, increases) for H ‖ to c. Fig. 7 shows the superconducting phase diagram elucidating the variation of $H_{c3}^{||c}(T)$, $H_{c2}^{||c}(T)$ and $H_{c2}^{||ab}(T)$ with reduced temperature ($T/T_c$). For extrapolating the data to 0 K, the model [41]

$$(4)\ H_{c2}(T) = H_{c2}(0)(1 - \left(\frac{T}{T_c}\right)^a)^b$$

is used (solid black lines in Fig. 7). The same model is used to fit $H_{c3}$. Using the extrapolation of the above-applied model, the upper critical fields are found to be $H_{c2}^{||c}(0)$ =4.47 T and $H_{c2}^{||ab}(0)$ =19.93 T. These values are in fair agreement with already reported values [42,43]. Using the Werthamer – Helfand - Hohenberg (WHH) formula $H_{c2}(0) = -0.69\, T_c (dH_{c2}/dT)_{T_c}$, the upper critical field values are estimated to be $H_{c2}^{||c}(0)$ =4.05 T and $H_{c2}^{||ab}(0) = 14.30$ T. This $H_{c2}^{||c}(0)$ is in fair agreement with both models, however the measured $H_{c2}^{||ab}(0)$ differ. It signifies the superconductivity beyond the BCS regime in the type-II superconductor $NbSe_2$. To get a rough idea about the superconducting parameters, Ginzburg-Landau (GL) formula for coherence length is used, where the upper critical field given by $H_{c2}(0) = \frac{\Phi_0}{2\pi\xi_{GL}(0)^2}$ , $\Phi_0 = h/2e$ is the flux quantum and $\xi_{GL}(0)$ is the G-L coherence length at T= 0 K. As a result, the G-L coherence length $\xi_{GL}(0)$ is found to be 9.01 nm and 4.8 nm, for H ‖ c and H ‖ ab, respectively. For 2D superconductivity, the orbital effects are weaker and the upper critical field is governed by the Pauli paramagnetic effect ($H_p$ in tesla) which is limited by the Chandrasekhar-Clogston (or Pauli) paramagnetic limit, $H_p \equiv 1.86 T_c$ at T=0 K [44]. For our $NbSe_2$, for Tc=7.3 K we obtain $H_p$ = 13.58 T (the dashed line in Fig.7). Note $H_{c2}^{||ab}(0)$ and $H_{c3}^{||c}(0)$ crosses the Pauli limit, whereas $H_{c2}^{||c}(0)$ is well within its limit. The 2D anisotropy can be determined by the ratio of the upper critical field at different orientations as given by $H_{c2}^{||c}(0)/ H_{c2}^{||ab}(0)$[45], the ratio of upper critical fields at different magnetic field orientations is found to be 0.28, which confirms the high anisotropic properties being present in $NbSe_2$.

Further, the anisotropic superconducting properties are elucidated through DC isothermal magnetization M-H and M-T measurements in both orientations. Figs. 8(a) and 8(c) show the ZFC and FC branches for (H ‖ to *c*) and (H ‖ to b) at 30 and 25.4 Oe respectively. A strong diamagnetic signal is observed at $T_c$=7.1(1) K for FC and ZFC plots in both orientations, attributed to a typical type-II superconductor [46]. Significant differences between ZFC and FC data suggest large pinning. The inset of Fig. 8(a) shows an enlarged view close to $T_c$. where the two branches merge at the irreversible temperature $T_{irr}$=6.8(1) K. Within the uncertainty no reversible region is observed for H ‖ to ab. Anyhow a very small reversible region occurs in both orientations. The inset of Fig. 8(c) shows similarity in the normalized FC magnetization plotted for both orientations.

Those figures also demonstrate the anisotropy present in the sample. For out-of-plane (H ‖ to c) the ZFC magnetic susceptibility value is ~ -8 at 5 K but only ~ -1 for the in-plane (H ‖ to ab) direction. Qualitatively, two factors may affect this discrepancy. (i) A higher flux pinning for H ‖ to c (the SC currents induced are in the plane) and (ii) due to the flake shape of the crystal, the demagnetization factor of the plane is much higher. Also, the superconductors' ideal susceptibility is -1(perfect diamagnetism), which is exceeded by experimental values. This is due to a large demagnetization factor which presently, we cannot evaluate including the contribution of each factor to the anisotropy.

From the isothermal magnetization measurement (M-H) two superconducting parameters can be determined: (i) the Meissner state down to the lower critical field ($Hc_1$) and (ii) the mixed state up to the upper critical field ($Hc_2$). Figs. 8(b) and (d) show the isothermal magnetization M-H plot at 4.5 K for the two orientations. Note also the higher magnetization values for H ‖ to *c*. For the out-of-plane (H ‖ to *ab*) orientation (Fig. 8(d)), the straight-line deviates and defines $H_{c1}^{||ab}$~ 100 Oe (shown in an extended scale in the inset), whereas the extrapolation to H=0 yields $H_{c2}^{||ab}$~ 4400 Oe. This field agrees well with the above value deduced by resistivity studies shown in Figs. 6(b) and (7). For H ‖ to *c*, $H_{c1}^{||c}$ is ~18 Oe, thus the $H_{c1}^{||ab}$/ $H_{c1}^{||c}$ ratio is ~5. The high $H_{c2}^{||c}$ (see Fig. 5(d) was not determined.

Moreover, to understand the electronic properties of $NbSe_2$, theoretical calculations are performed using the DFT-based software Quantum Espresso. The inset of Fig. 9(a) shows the high-symmetric path chosen for band structure calculation in the first Brillouin zone (BZ). The optimized path for the band structure calculation is obtained from the work by Hinuma et. al. [47]. The band structure calculation is performed with different conditions for the Bulk 2H-$NbSe_2$, monolayer and bilayer. The calculated band structure of Bulk 2H-$NbSe_2$ is shown in Fig. 9(b) along the high symmetry path Γ-M-K-Γ-A-L-H-A [47]. To get accurate band structure several corrections and hybrid functional are developed such as B3LYP hybrid functional [48], MP2 [49] and DFT-D3 [50] for Van der Waals corrections which make computational simulation more realistic. In TMDs, atomic layers of X-M-X interact via the Van der Waals force, therefore, the DFT-D3 corrections are included in the structural optimization stage and at the band structure calculations. The dispersion correction energy due to Van der Waals interaction is found to be -0.0964 eV. Fig. 9(b) shows the band structure without SOC (w/o SOC) and with SOC parameters. Small band openings ~0.157 eV are observable at the high symmetry point K due to the SOC effect. The w/o SOC bands show a possible double degenerate type-II Dirac-cone along the path K-Γ at around 0.2 eV below the Fermi energy (see Fig. 9(c)). Even when the SOC effect is considered in the band calculation, the Dirac cone degeneracy remains intact. Dirac cone Type I or II is determined by the tilt of the cone. Because of the tilt of the Dirac cone in the band structure, $NbSe_2$ may belong to the type II Dirac semimetals according to preliminary DFT calculations. Along path A-L, the two bands crossing the Fermi level are doubly degenerate throughout, for both cases – w/o SOC band and SOC bands, indicating nodal line features present in the $NbSe_2$. These primary interesting results need further theoretical and experimental verification for which Angle-resolved photoemission spectroscopy (ARPES) is warranted. Fig. 9(d) shows the Bulk band structure calculated within the GGA+U framework. Below the Femi level, bands are minutely perturbed, but near the Fermi level, the band structure features are equivalent to that of GGA+SOC bands structure namely Dirac cone and nodal line are observable. Overall, in the Bulk band structure, three bands cross the Fermi level, which is in support of the observed metallic behavior in resistivity measurement. The DFT magnetization calculations under both conditions (with SOC effect and Hubbard correction U) show that the Bulk $NbSe_2$ is non-magnetic with total magnetization 0 $\mu_B$, which is well supported by the magnetization measurement above $T_c$.

Further, Fig. 9(e) shows the band structure calculated for the monolayer (ML) of $NbSe_2$ for un-relaxed and relaxed structures. Non-relaxed mono-layer is directly created from XRD-fitted data and then relaxed within DFT limits. Since, it is a single layer, no DFT-D3 correction is implemented as Van der Waals interaction is absent. The non-relaxed band structure implies a direct band gap of 1.47 eV. Interestingly, the relaxed structure shows the presence of a single flat band crossing the Fermi level which is also reported by [51]. The magnetization calculation shows that ML has a ferromagnetic character with a total magnetization of ~1 $\mu_B$ per formula unit. When the Hubbard correction is included with the relaxed structure, the electronic band gap is observed. Fig. 9(f) shows the spin-resolved band structure with Hubbard correction. The spin-up band shows a band gap of 1.38 eV at high symmetry point K whereas, the spin-down band has an indirect band gap of 1.17 eV. Thus, $NbSe_2$ monolayer is expected to show insulator/semiconductor behavior.

Fig. 9(g) shows the band structure for the bi-layer of $NbSe_2$, for three cases of un-relaxed structure, relaxed and relaxed with Hubbard correction. There are two bands across the Fermi level implying conducting behavior in all three cases without significant differences. A Dirac cone similar to the Bulk band in the vicinity of the Fermi level along path K-Γ is observable. Also, the DFT calculation shows that the bilayer is non-magnetic. Thus, the bilayer almost starts resembling the Bulk behavior of $NbSe_2$. An insulator-to-metal transition is evident from monolayer to bilayer. The transition may have two possible origins (i) The structural change due to the addition of another layer (Van der Waals interaction) and (ii) the Antiferromagnetic coupling of the two layers. The monolayer and bilayer have different magnetic phases. The results observed by DFT calculations are summarized in Table 1. Note that in the case of mono-layer, the Van der Waals interaction is not present, since there is no other layer of nearby $NbSe_2$. The effect of DFT-D3 correction simulating Van der Waals interaction is illustrated through bilayer relaxation structure calculation as shown in Table 1. A change in Nb-Se bond length, Se-Nb-Se bond angle and Nb-Nb separation of the two layers is significantly evident when DFT-D3 correction is implemented in the structure relaxation calculation. The DFT calculations approve the experimental results exhibiting bulk non-magnetic metallic behavior. The spin-orbit coupling is also effective in the studied material, showing the possibility of the presence of topological character. Further, theoretical and experimental studies are being carried out to find out the true topological nature of $NbSe_2$.

## Conclusion

In summary, the article addresses the anisotropy in the electronic and magneto-transport of CVT grown 2D superconductor $NbSe_2$. The XRD (X-ray Diffractometry) of as grown mechanically cleaved crystal elucidated that the crystal is grown uni-directionally in a single phase and is well-fitted in a hexagonal structure. The stoichiometric ratio of elements in as-grown $NbSe_2$ is found to be Nb:Se (1.01:1.99). The recorded Raman spectra illustrated four Raman active vibrational modes as predicted for the TMDs. The superconducting transition is observed at below 7.3 K. The ρ-H behaviour of as-grown crystal elucidated a high critical field in both in-plane and out-of-plane measurements, albeit with anisotropy being present in the sample. For H || to c, the magneto-transport data presumably hints towards the existence of the filamentary state. The upper critical field for in-plane orientation crosses the Pauli paramagnetic limit. The magnetization in ZC and ZFC protocol is observed at 30 Oe, elucidating the transition temperature at 7.1 K with an irreversible temperature of 6.8 K for the field parallel to the c axis. The diamagnetic behaviour of the M-H curve resembles a typical type-II superconductor. The density functional theory (DFT) confirmed the band structure of $NbSe_2$ to be metallic but the same for mono-layers showed a band gap of 1.17 eV.

## Acknowledgment

The authors would like to thank the Director of the National Physical Laboratory (NPL), India, for his keen interest in the present work. The authors acknowledge Dr. J. Tawale and Ms. Sweta for SEM/EDAX and Raman spectroscopy measurements, respectively. Also, N.K. Karn would like to thank CSIR, India, for the research fellowship, and AcSIR-NPL for Ph.D. registration. This work is supported by in-house project numbers OLP 240832 and OLP 240232.

## Conflict of Interest Statement

The authors have no conflict of interest.

Table 1

Summary of DFT calcualtions

| | | Bulk | ML | BL |
|---|---|---|---|---|
| NbSe2 System | Exp. Cell | Nb-Nb 6.277 Å<br>Nb-Se 2.531 Å<br>Se-Nb-Se 76.35$^0$ | | |
| | Relaxed Cell | | Nb-Se 2.596 Å<br>Se-Nb-Se 79.93$^0$ | Nb-Nb 6.164 Å<br>Nb-Se 2.587 Å<br>Se-Nb-Se 79.89$^0$ |
| | Relaxed Cell DFT-D3 | Nb-Nb 6.175 Å<br>Nb-Se 2.591 Å<br>Se-Nb-Se 80.43$^0$ | NA | Nb-Nb 5.942 Å<br>Nb-Se 2.578 Å<br>Se-Nb-Se 79.67$^0$ |
| Magnetic property | | Non-magnetic | Magnetic<br>$M_{tot}$ ~1 $\mu_B$ | Non-magnetic |
| GGA (Not relaxed) | | Metallic | Insulator | Metallic |
| GGA (Relaxed) | | Metallic | Metallic | Metallic |
| GGA+U | | Metallic | Insulator | Metallic |

## Figure Captions

**Figure 1:** **(a)** XRD pattern taken on a thin crystal flake of synthesized $NbSe_2$ single crystal. The inset shows the unit cell generated by VESTA. **(b)** Rietveld refinement of PXRD pattern of $NbSe_2$.

**Figure 2:** **(a)** SEM image of surface morphology of synthesized $NbSe_2$ single crystal. **(b)** EDAX spectra of synthesized $NbSe_2$ single crystal, in which the inset table shows the elemental composition of constituent elements of $NbSe_2$.

**Figure 3:** Lorentz deconvoluted room temperature Raman spectra of $NbSe_2$.

**Figure 4:** Temperature-dependent resistance of $NbSe_2$ at zero field and the inset shows the enlarged image near the transition temperature.

**Figure 5:** For H||c field orientation, **(a), (b)** and **(c)** show the temperature-dependent resistivity of $NbSe_2$ for various applied fields. Inset of (a) depicts the experimental setup, sample and field orientation. **(d)**The magnetic field dependence of the resistivity ρ(H) in $NbSe_2$ at temperatures 2 K, 4 K, 6 K and 8 K. The MR% at 8 K is shown in the inset.

**Figure 6:** For H||ab field orientation, **(a)** the temperature-dependent resistivity of $NbSe_2$ for various applied fields. **(b)**The magnetic field dependence of the resistivity ρ(H) in $NbSe_2$ at temperatures 2 K, 4 K, 6 K and 8 K.

**Figure 7:** Variation of the critical field with reduced temperature for in-plane and out-of-plane applied fields.

**Figure 8:** **(a)** DC magnetization measurements of synthesized $NbSe_2$ crystal at magnetic field 30 Oe parallel to c axis under FC and ZFC protocols and the inset shows enlarged image near $T_c$. **(b)** Isothermal magnetization at 4.5 K with magnetic field along c-axis. **(c)** DC magnetization at magnetic field 25.04 Oe in ab plane under FC and ZFC protocol **(d)** Isothermal magnetization at 4.5 K with magnetic field in ab plane.

**Figure 9:** **(a)** The k-path in the first Brillouin zone guided by green arrow. **(b)** The band structure of bulk $NbSe_2$ for both SOC and without SOC parameters. **(c)** The enlarged view of the band structure in the path K-Γ. **(d)** Bulk band structure with Hubbard correction and the inset shows an enlarged view in path K-Γ. **(e)** and **(f)** shows the band structure of the mono-layer with un-relaxed, relaxed and spin-polarised Hubbard correction implemented band structure. (g) Band structure of the bi-layer of $NbSe_2$ with un-relaxed, relaxed, and spin-polarised Hubbard correction implemented band structure.

**Figures:**

Fig. 1(a)

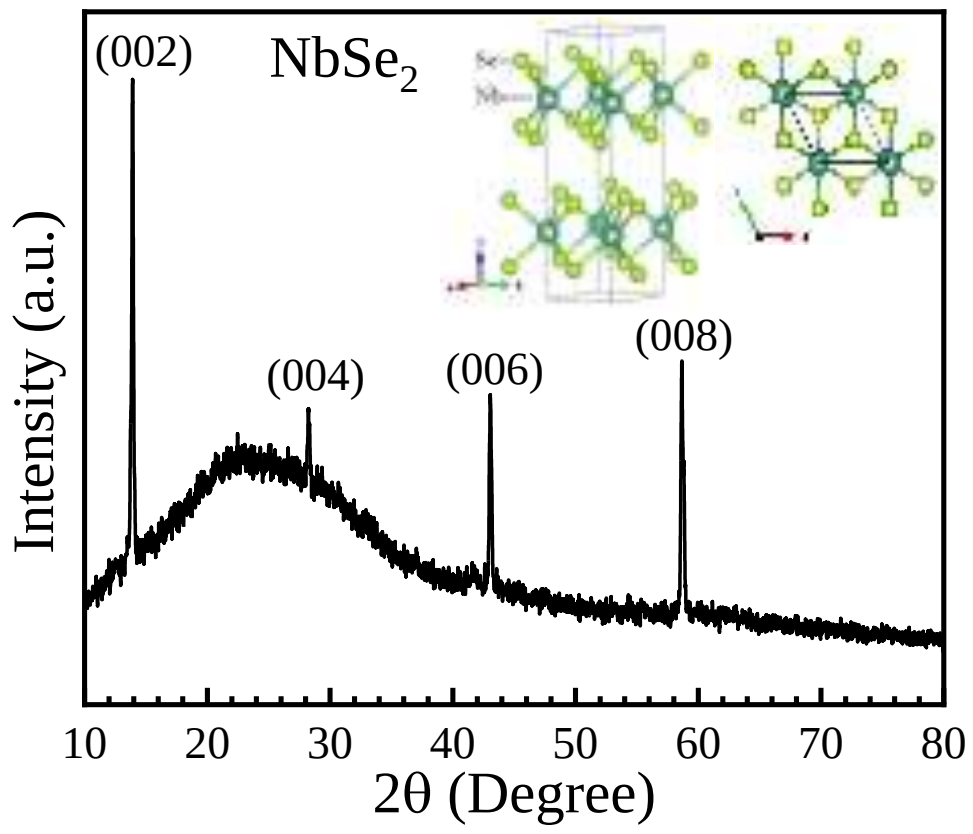

Fig. 1(b)

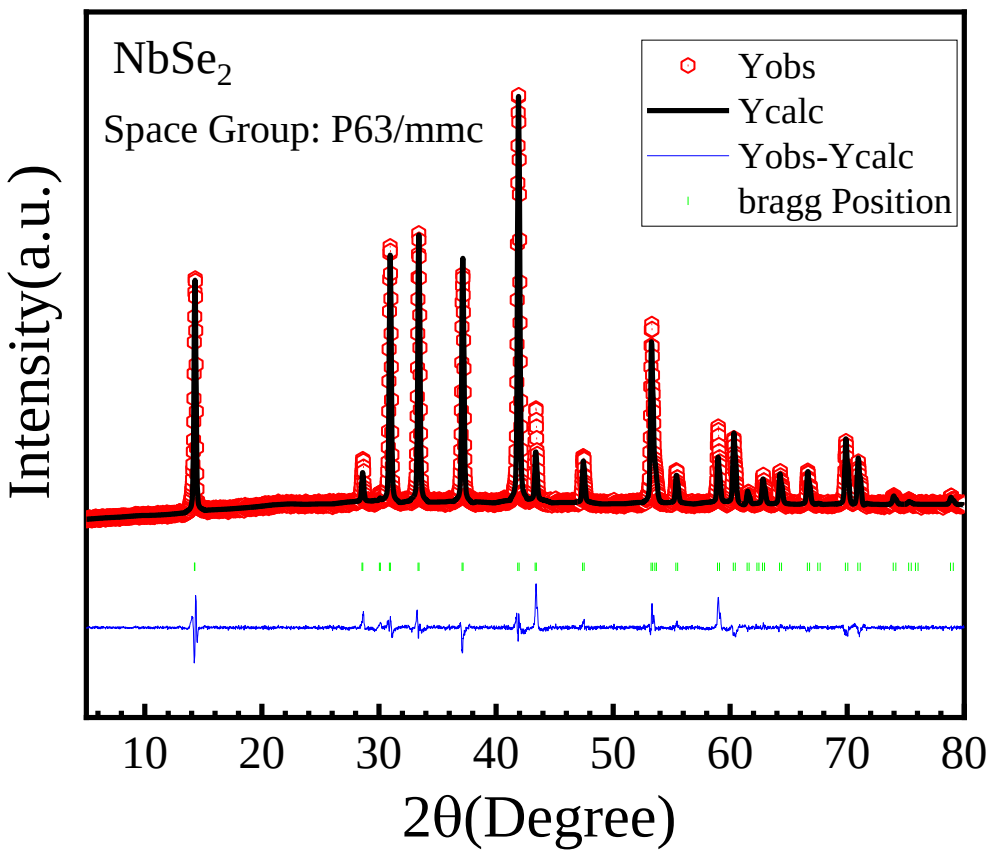

Fig. 2(a)

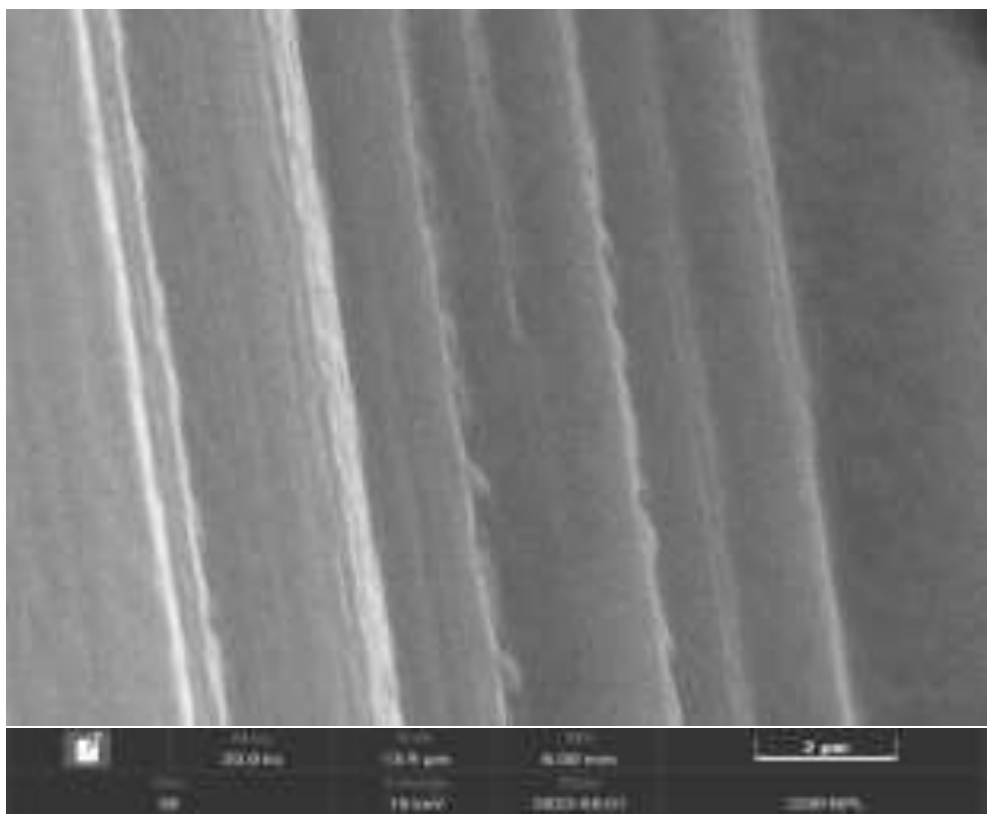

Fig. 2(b)

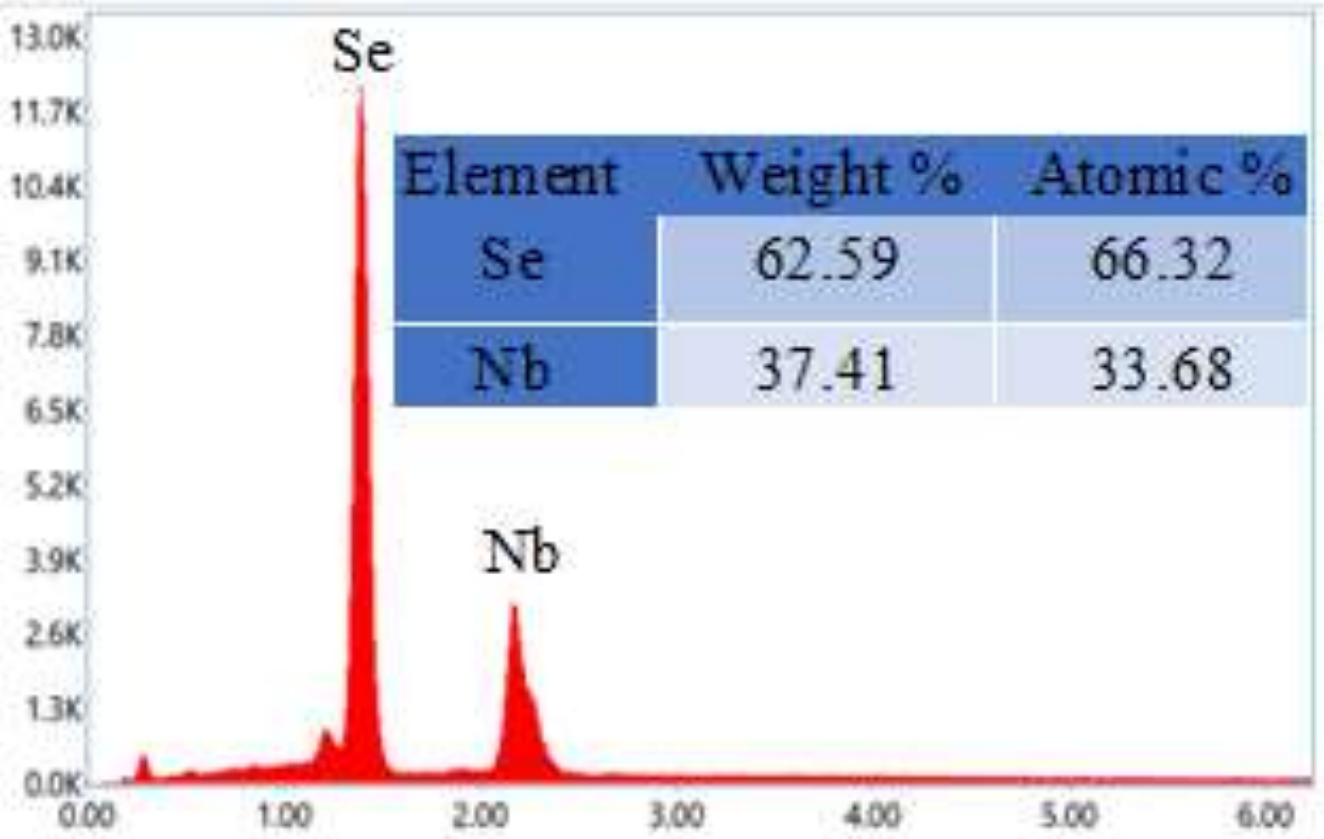

| Element | Weight % | Atomic % |
|---|---|---|
| Se | 62.59 | 66.32 |
| Nb | 37.41 | 33.68 |

Fig. 3

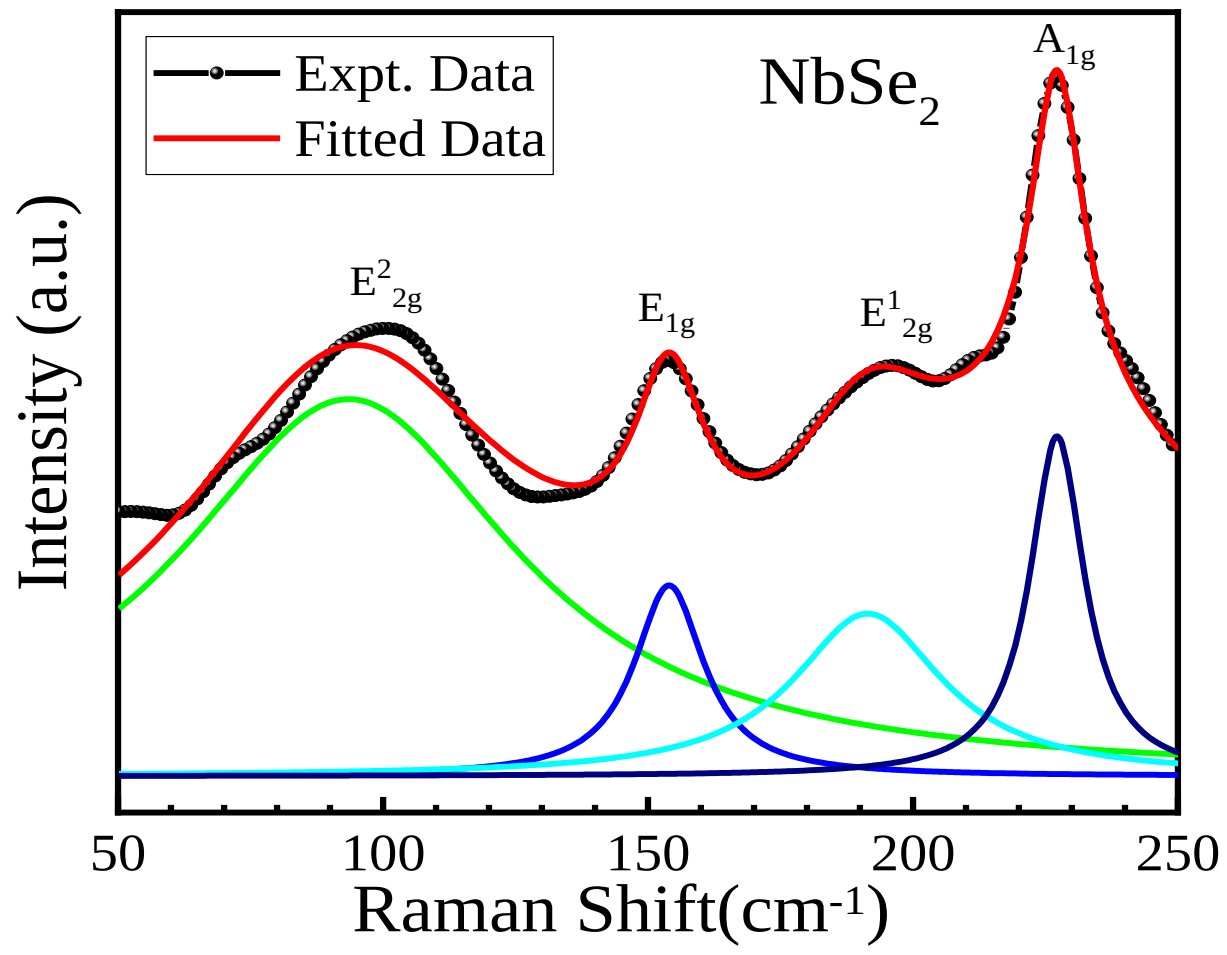

Fig. 4

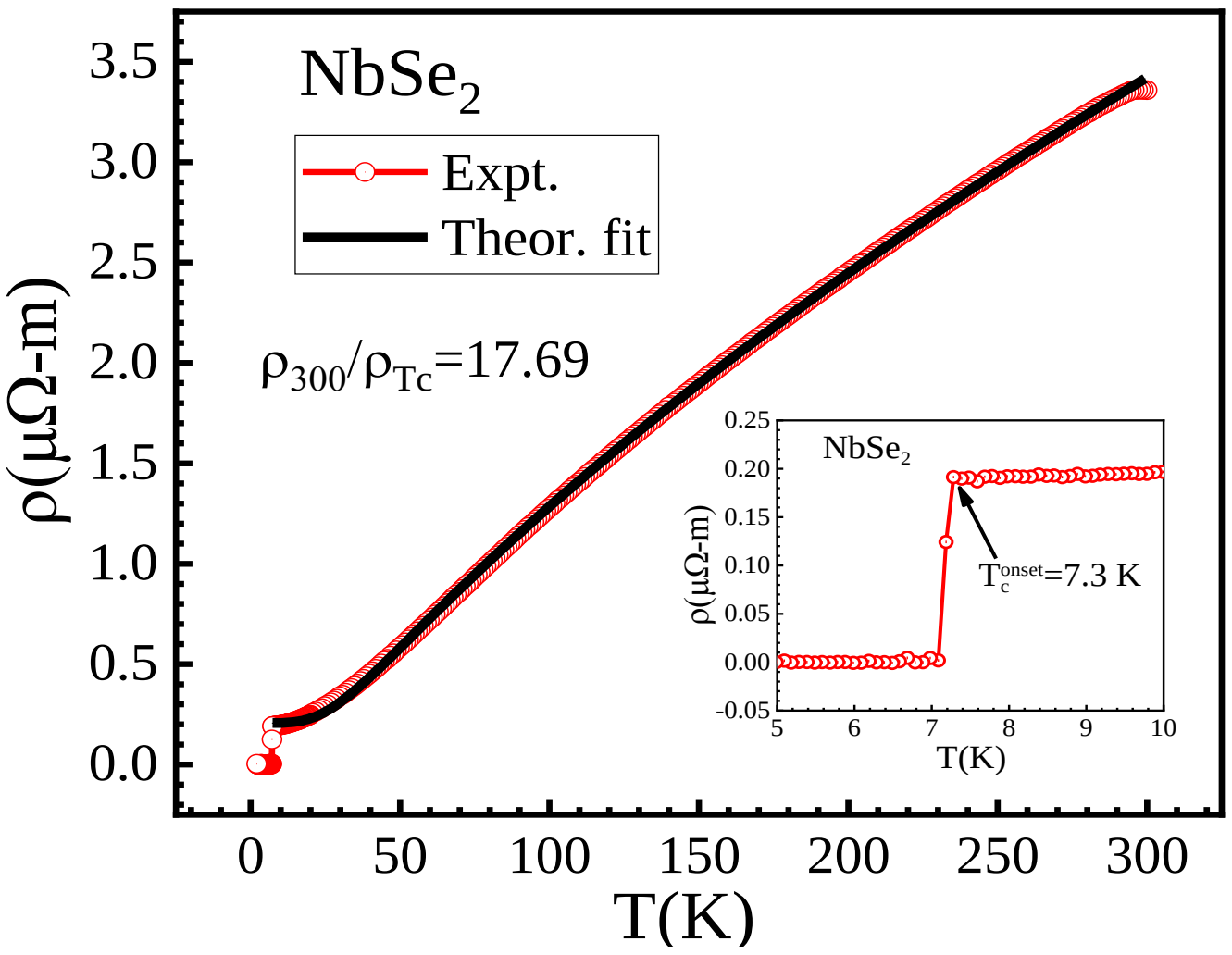

Fig. 5(a)

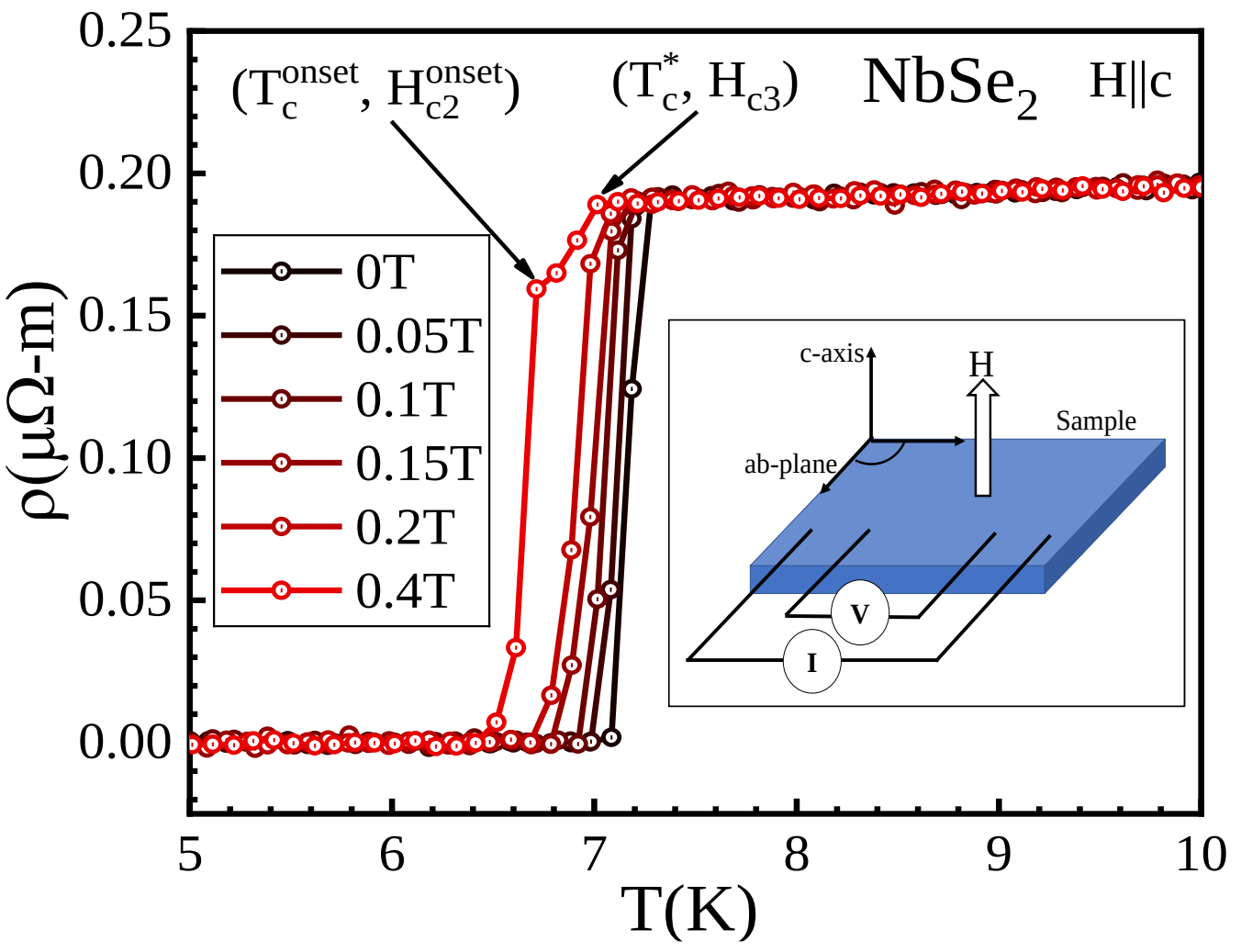

Fig. 5(b)

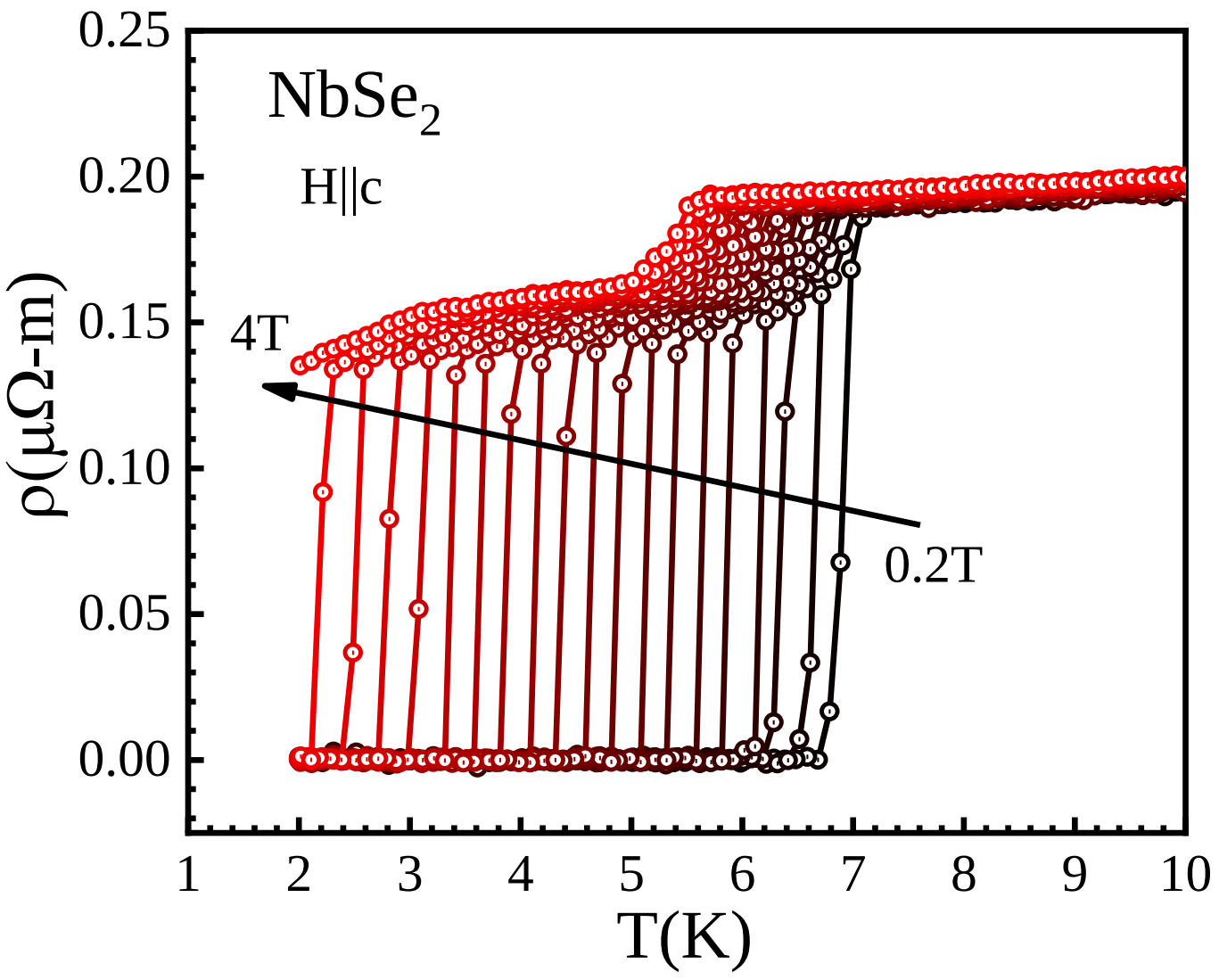

Fig. 5(c)

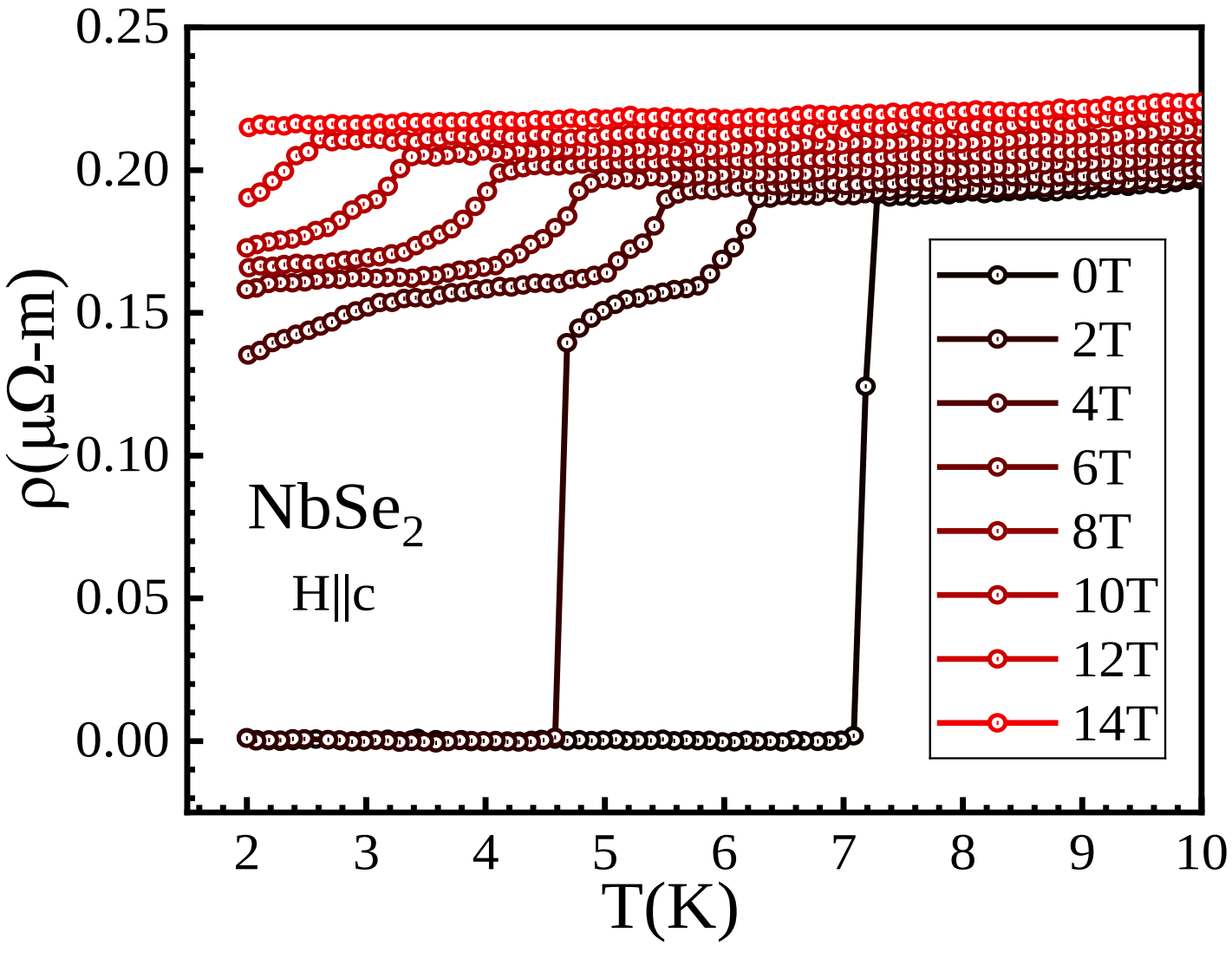

Fig. 5(d)

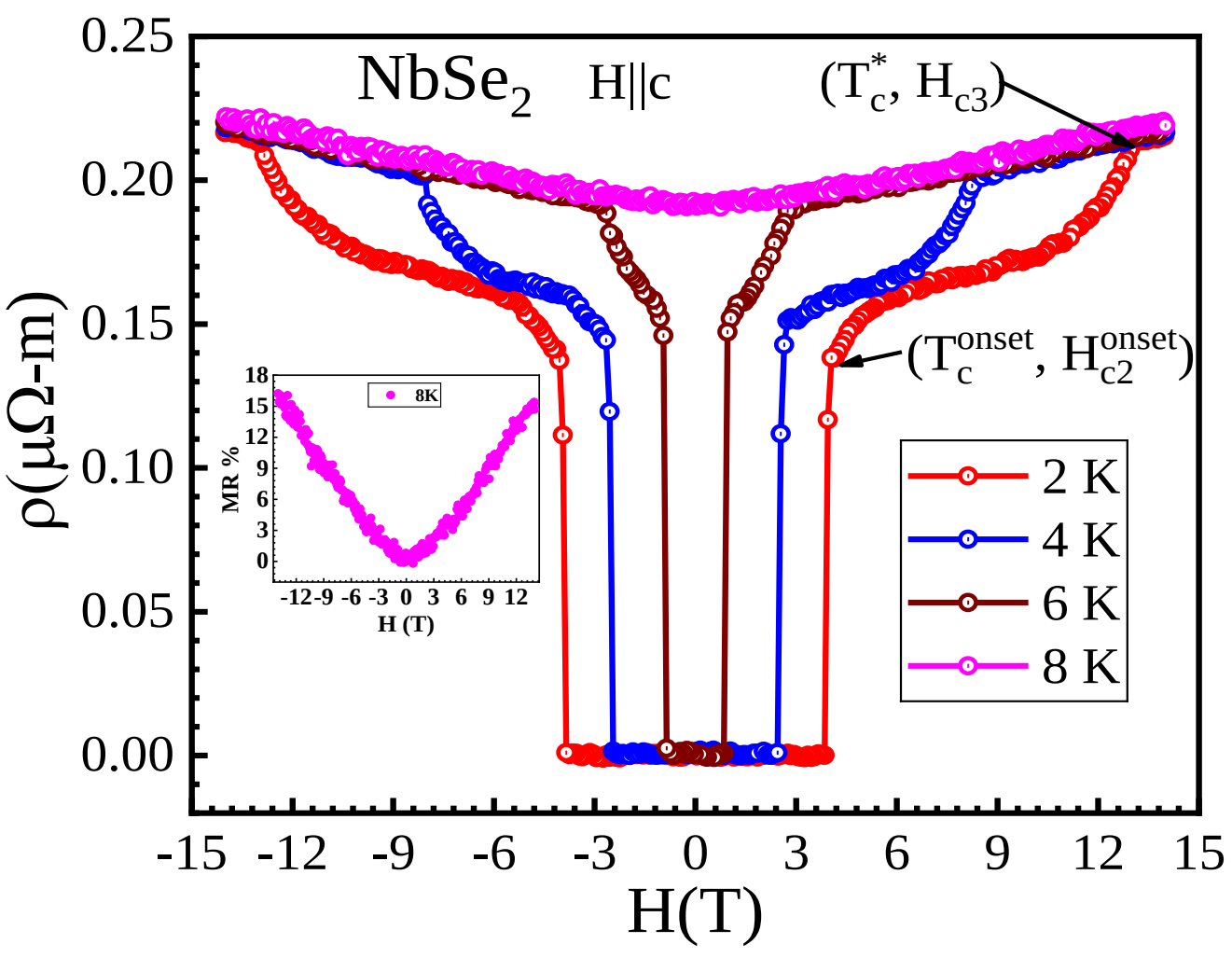

Fig. 6(a)

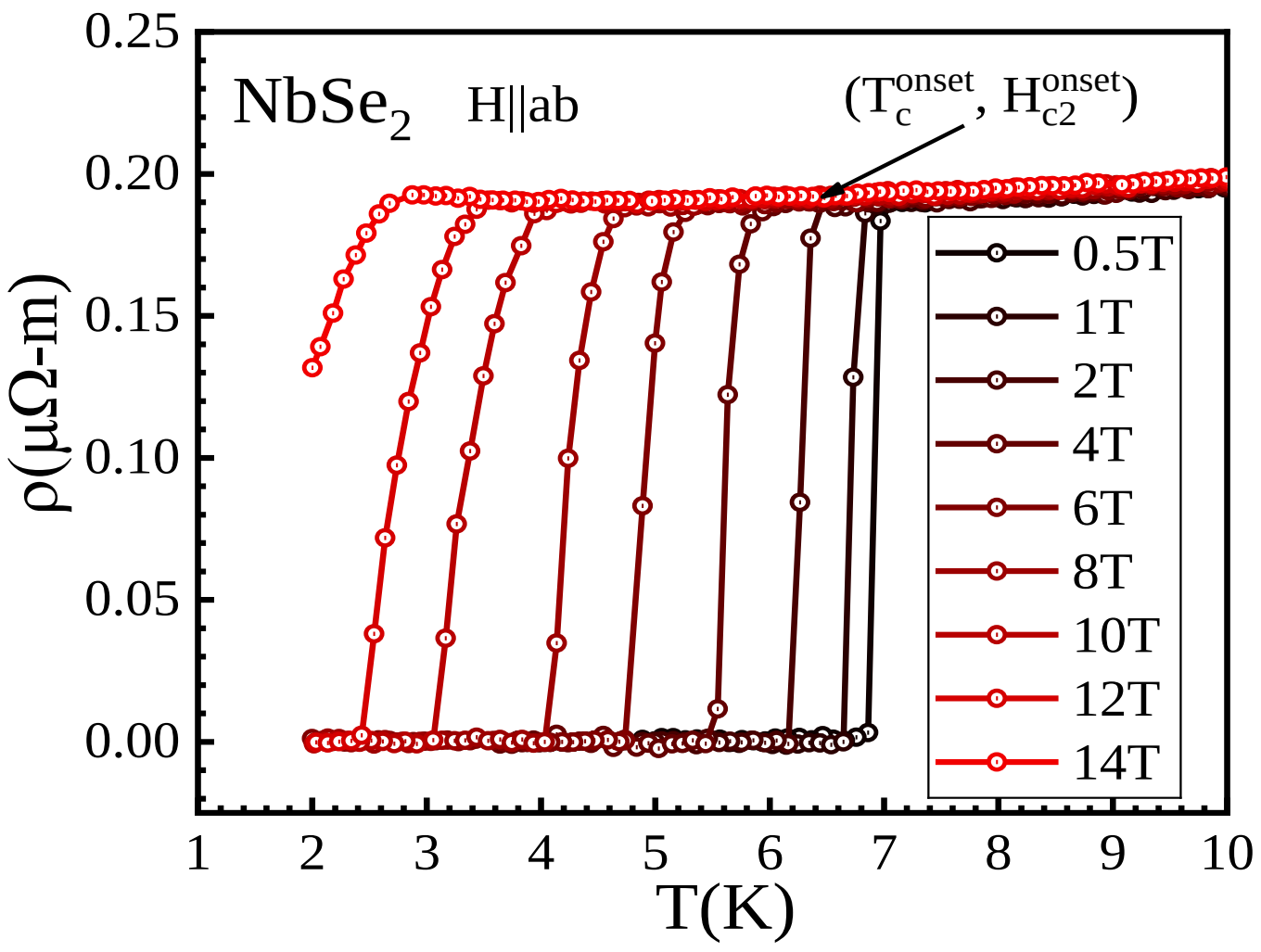

Fig. 6(b)

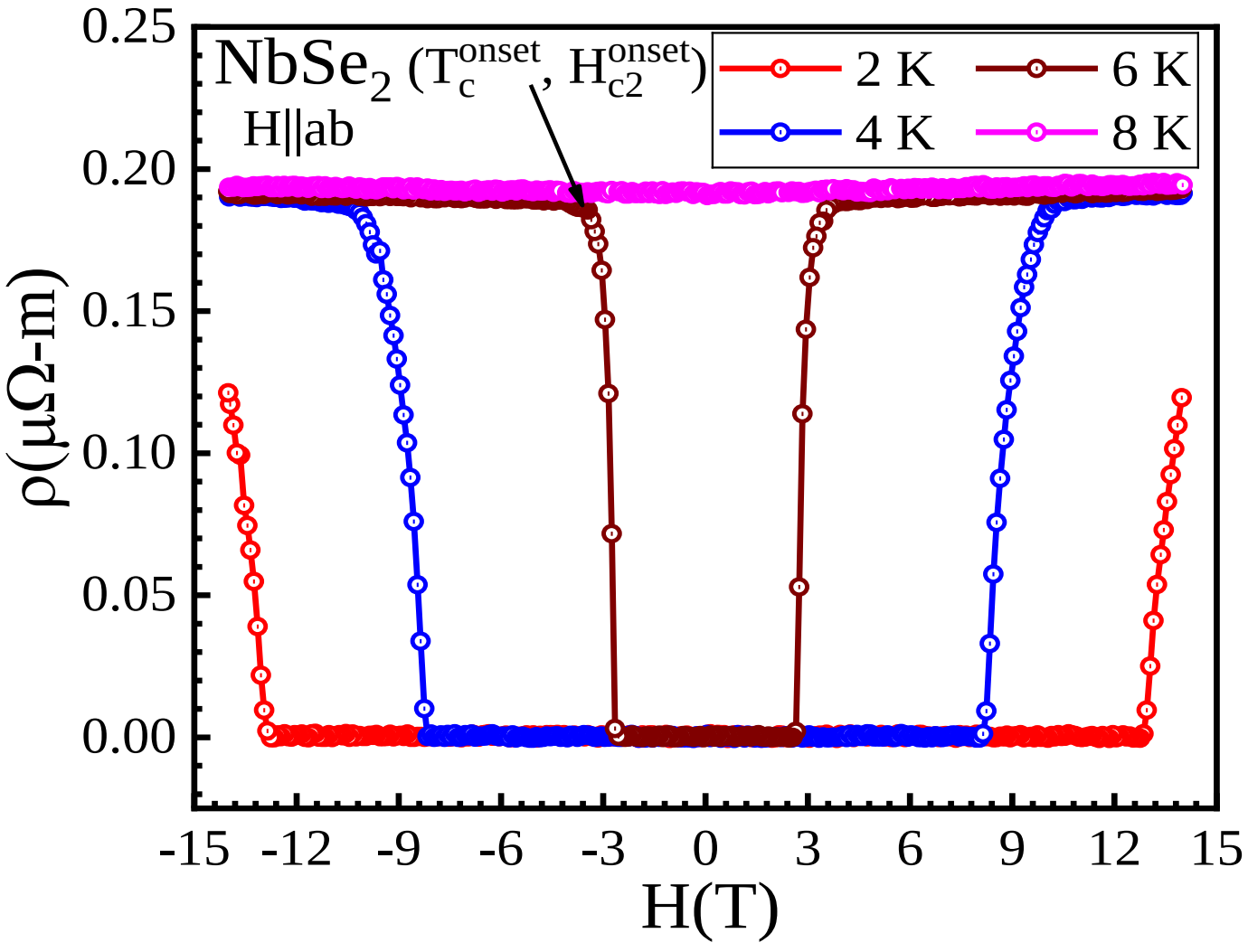

Fig. 7

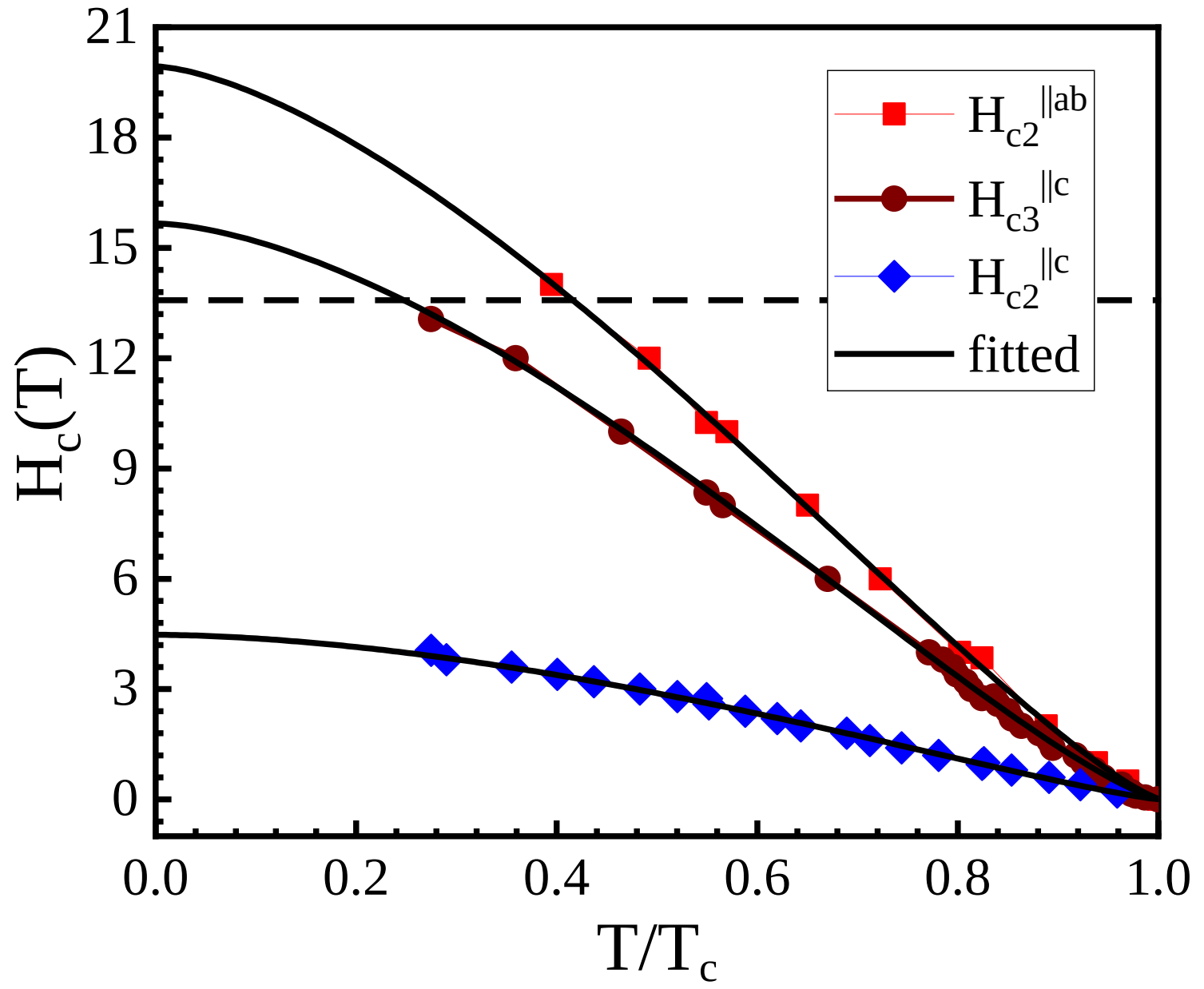

Fig. 8(a)

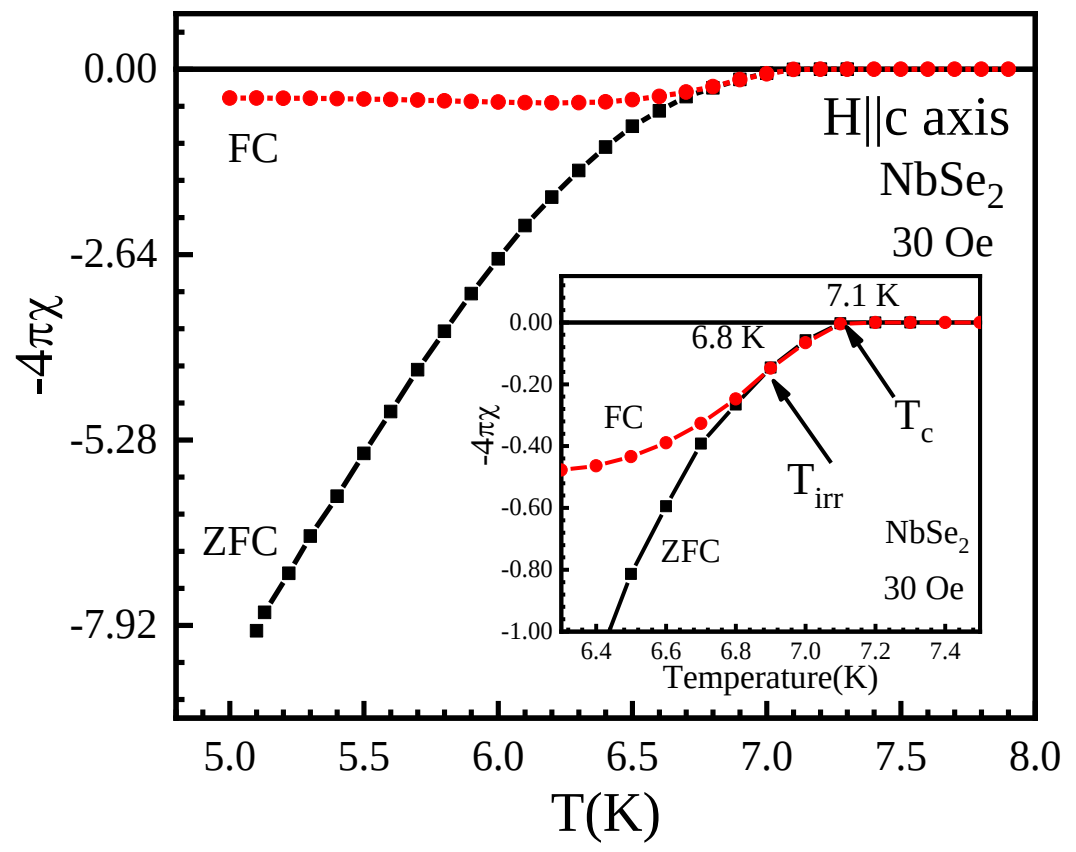

Fig. 8(b)

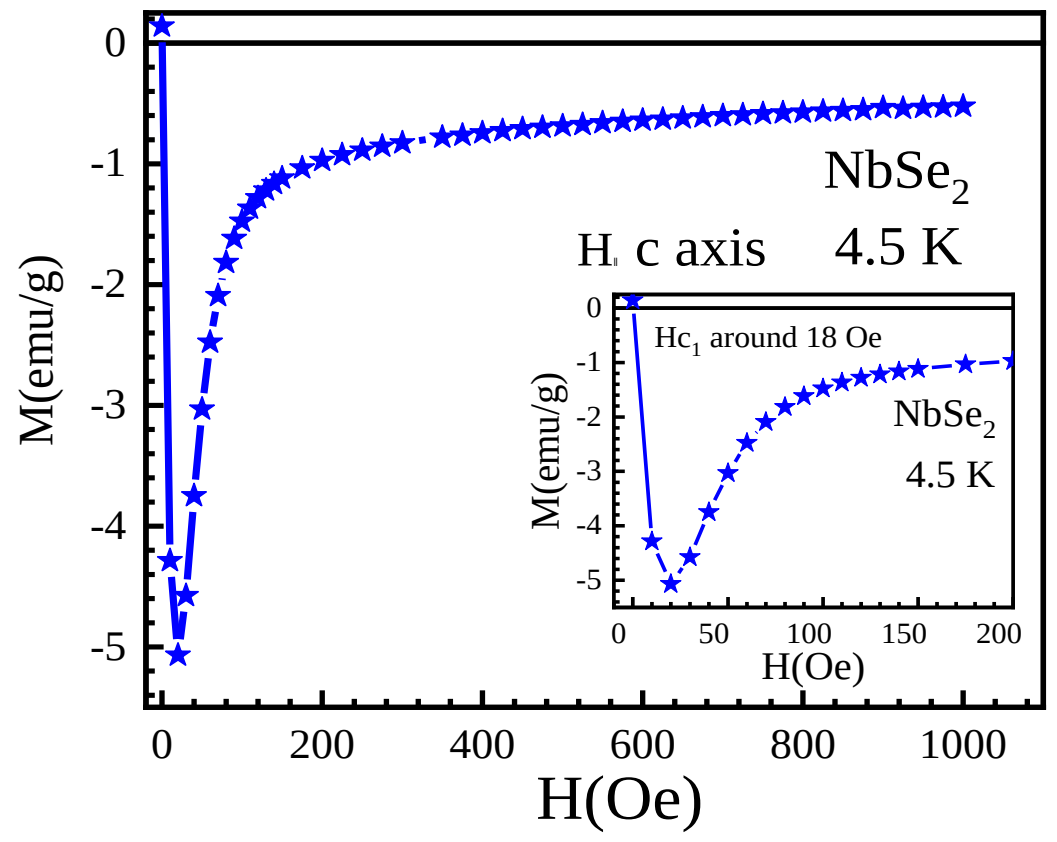

Fig. 8(c)

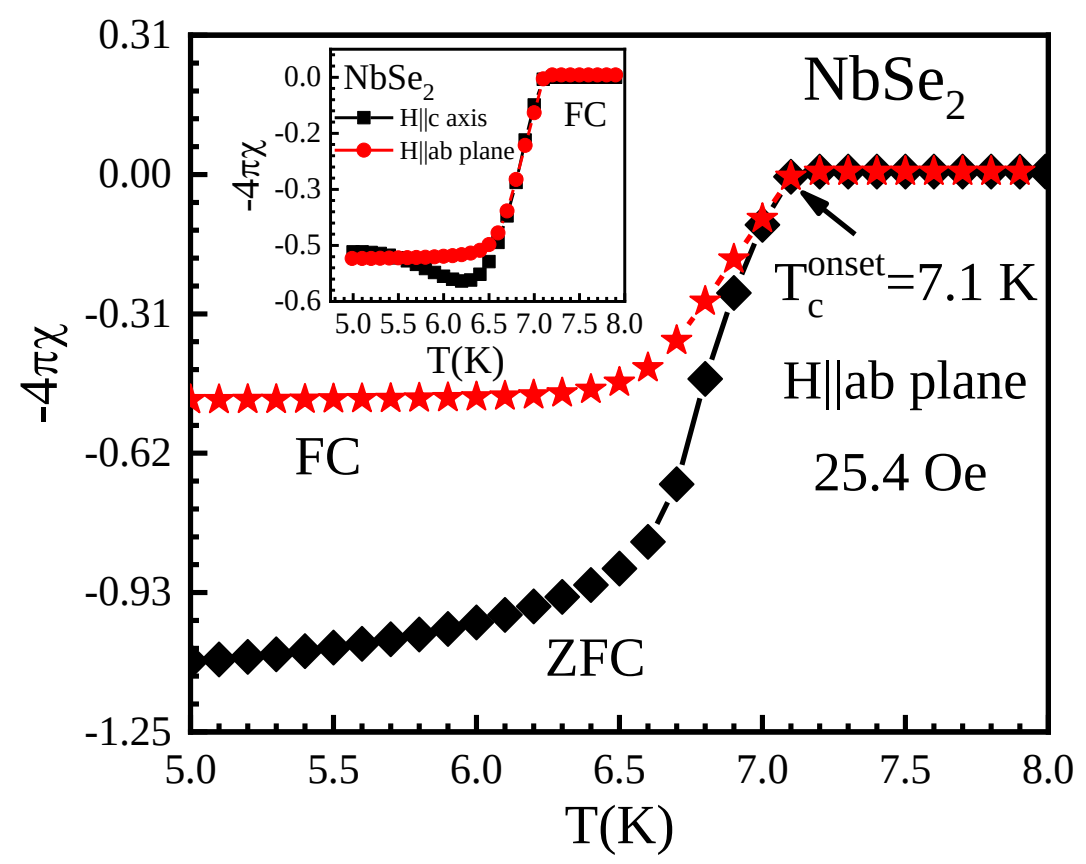

Fig. 8(d)

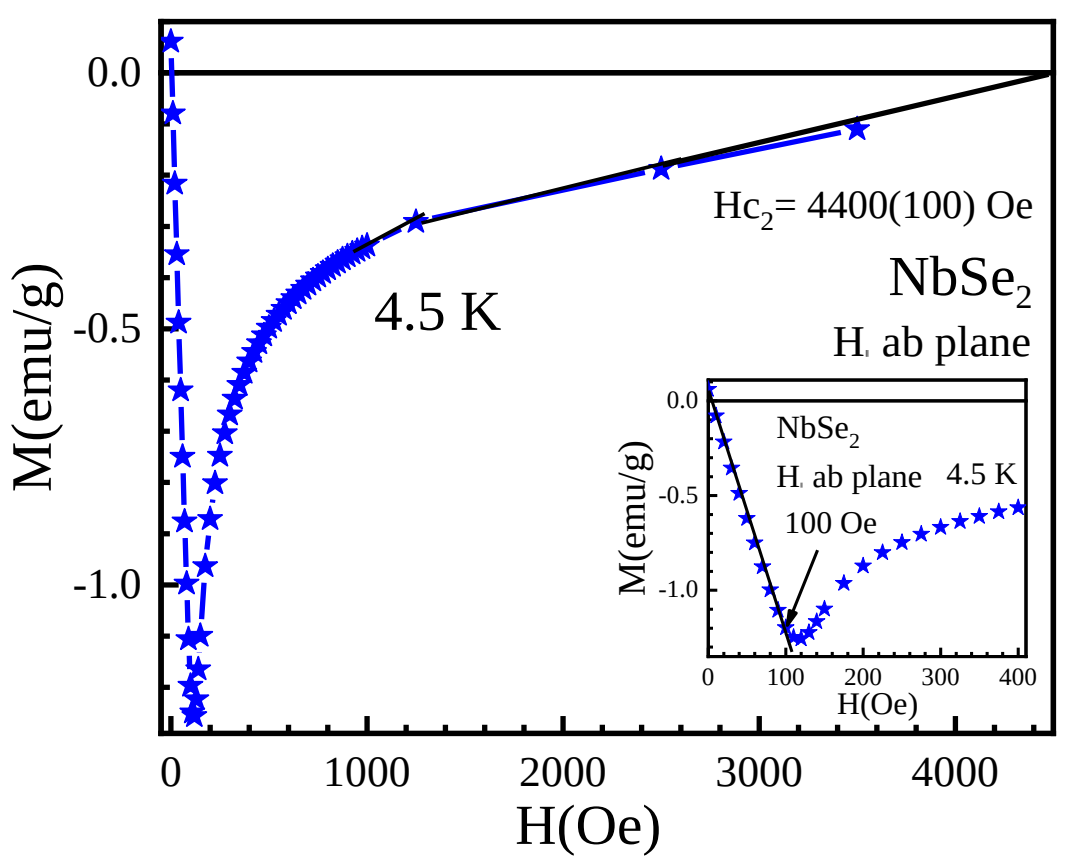

Fig. 9(a)

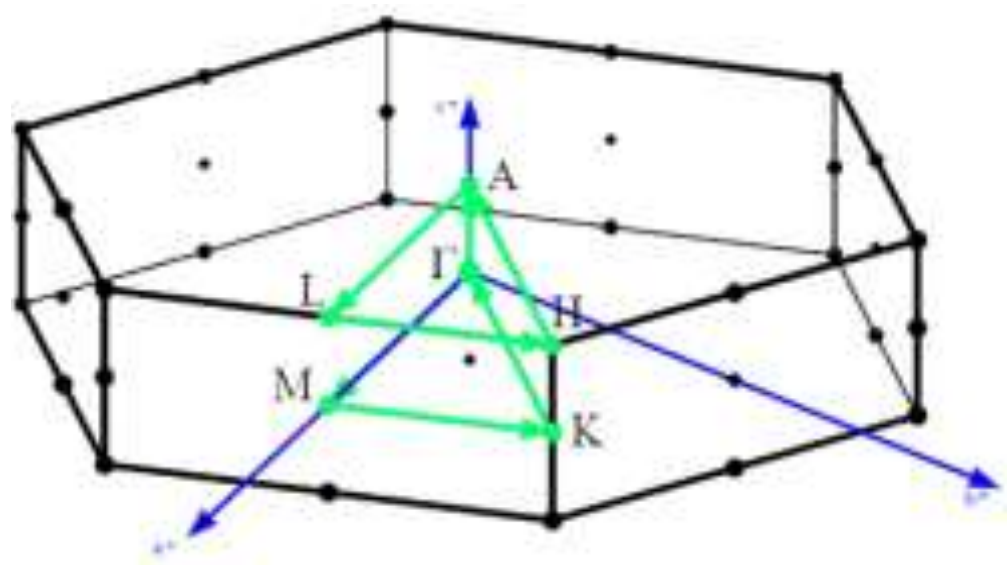

Fig. 9(b)

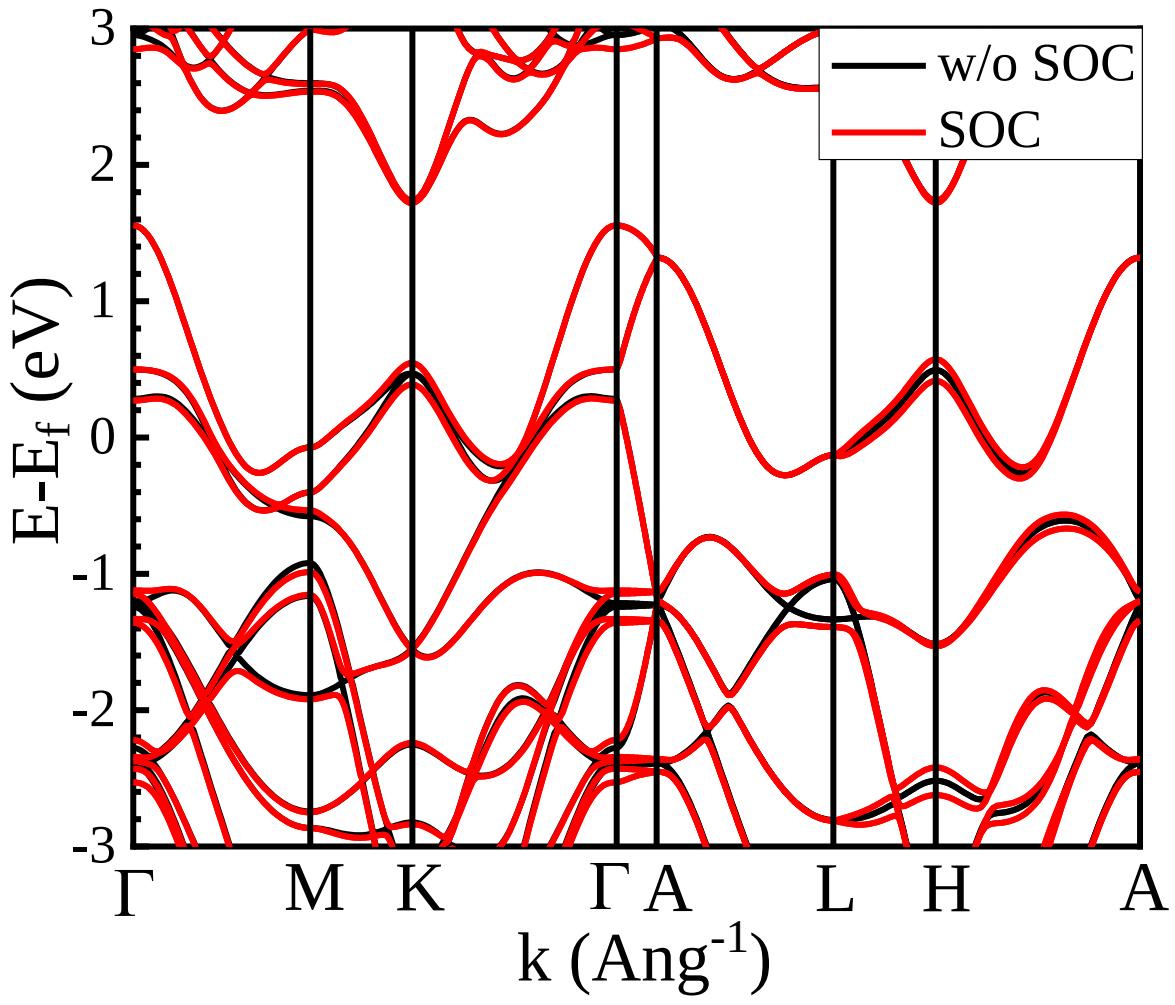

Fig. 9(c)

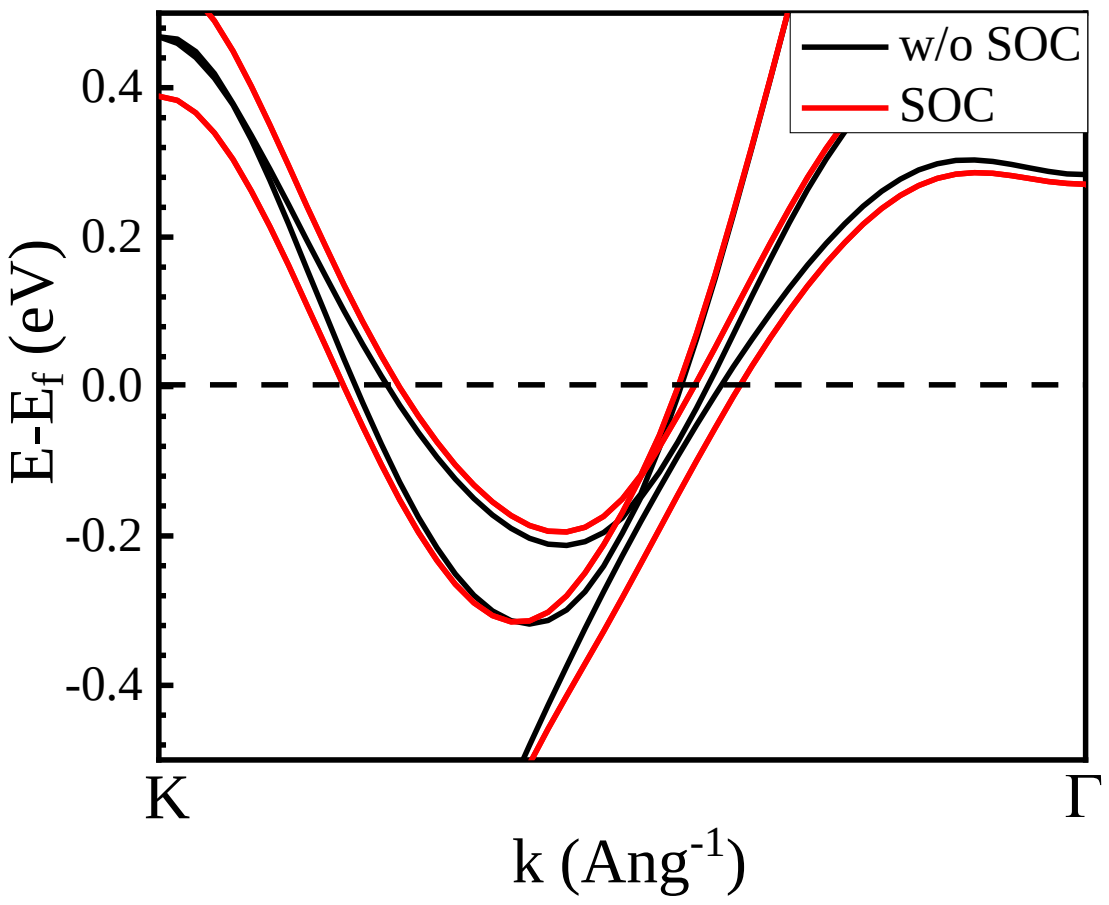

Fig. 9(d)

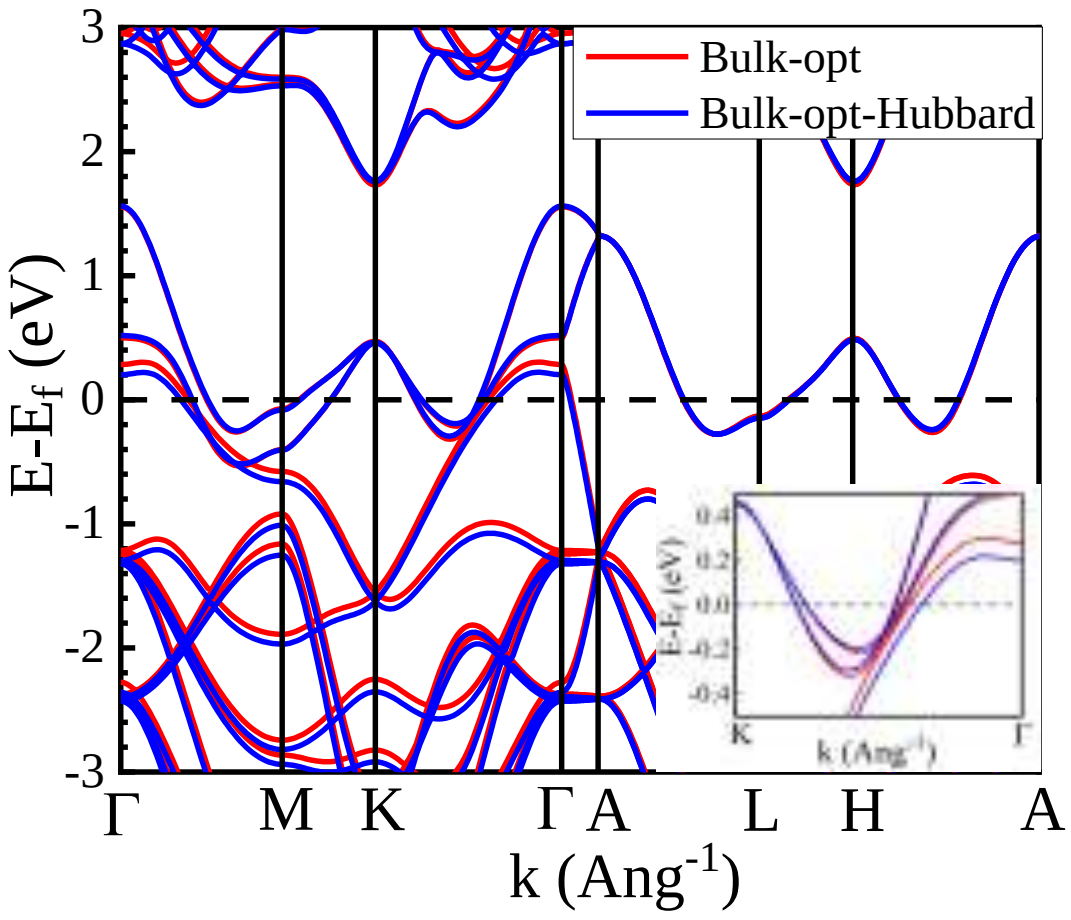

Fig. 9(e)

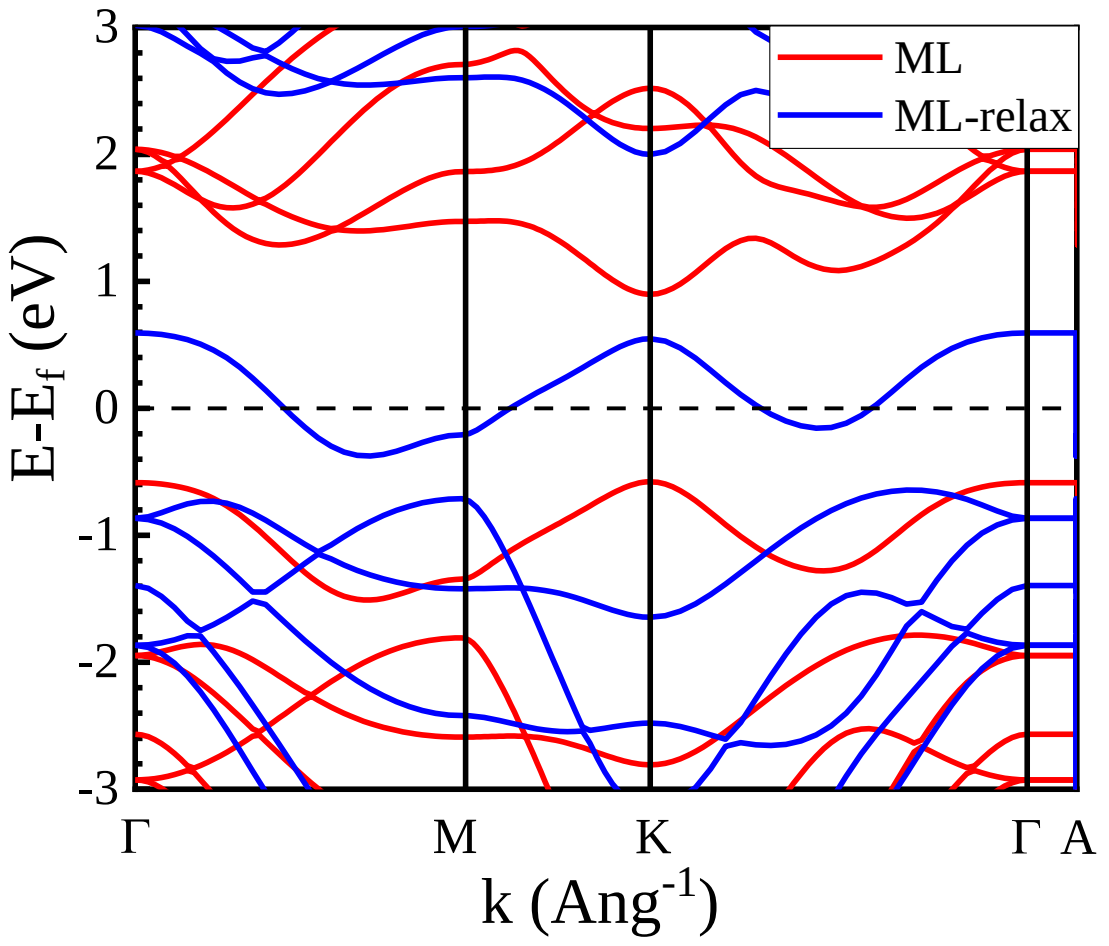

Fig. 9(f)

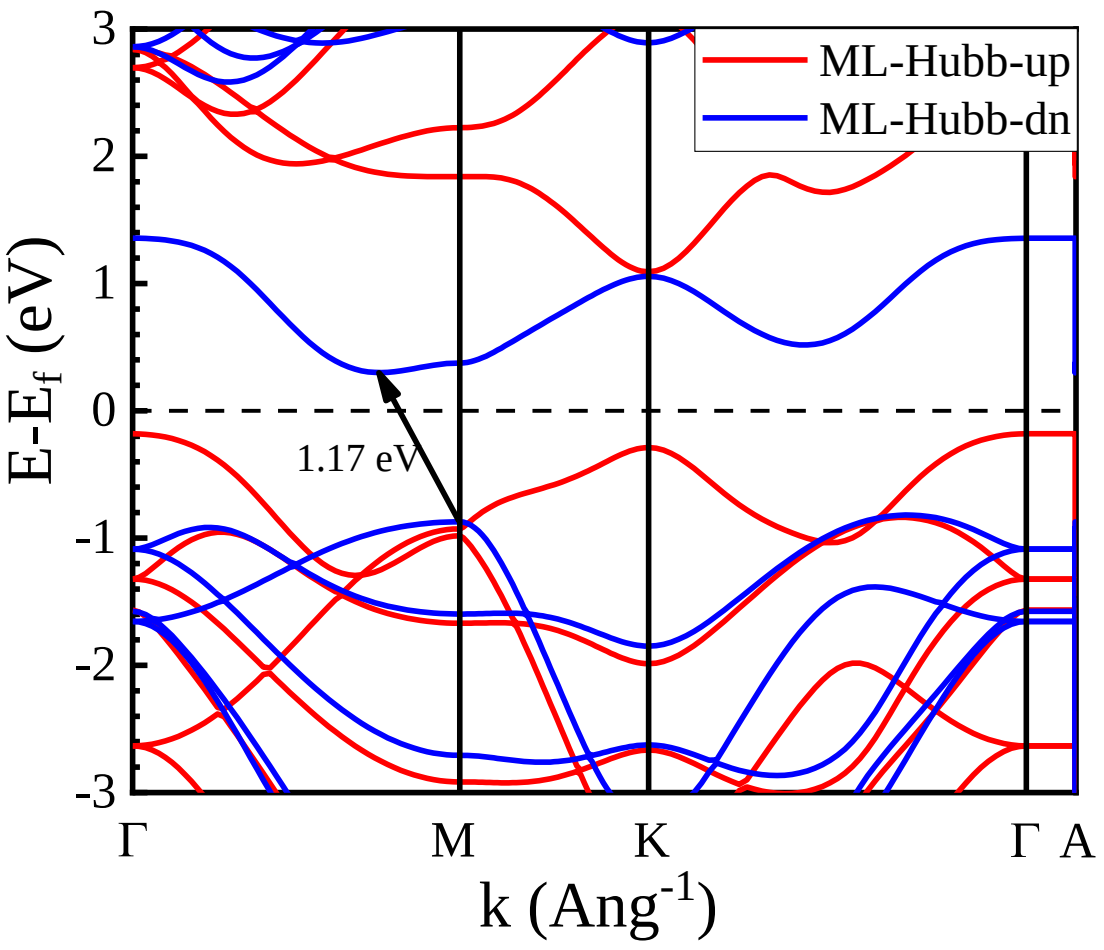

Fig. 9(g)

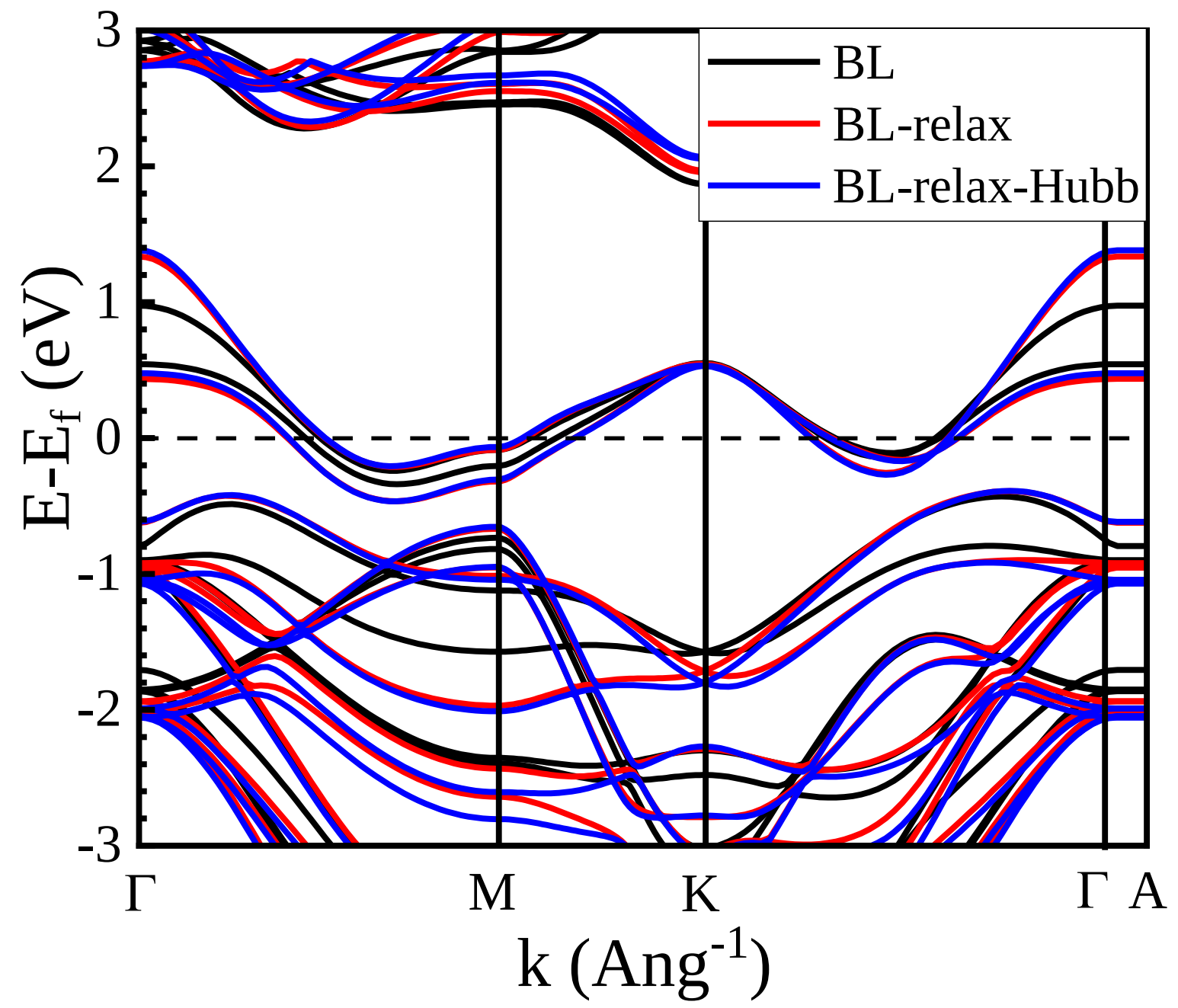